\documentclass[10pt, conference, compsocconf]{IEEEtran}

\usepackage{cite}
\usepackage{amsmath,amssymb,amsfonts}
\usepackage{algorithmic}
\usepackage{graphicx}
\usepackage{textcomp}
\usepackage{xcolor}
\usepackage[hyphens]{url}
\usepackage[ruled,vlined]{algorithm2e}
\usepackage{multirow}
\usepackage{soul}

\def\BibTeX{{\rm B\kern-.05em{\sc i\kern-.025em b}\kern-.08em
    T\kern-.1667em\lower.7ex\hbox{E}\kern-.125emX}}

\title{Hercules: Heterogeneity-Aware Inference Serving for At-Scale \\ Personalized Recommendation}

\author{
\normalsize 
Liu Ke$^\ast{}^\dagger$,
Udit Gupta$^\ast{}^\ddagger$,
Mark Hempstead$^\diamond$,
Carole-Jean Wu$^\ast$, 
Hsien-Hsin S. Lee$^\ast$, 
Xuan Zhang$^\dagger$ \\
\\
\normalsize $^\ast$Meta, $^\dagger$Washington University in St. Louis, $^\ddagger$Harvard University, $^\diamond$Tufts University
} 

\newif\ifrev
\ifrev
  
  \newcommand{\cjw}[1]{{\color{magenta} #1}}
  \newcommand{\xuan}[1]{{\color{red} [Xuan: #1]}}
  \newcommand{\sean}[1]{{\color{orange} [Sean: #1]}}
  \newcommand{\todo}[1]{{\color{cyan} [TODO: #1]}}
  \newcommand{\comment}[1]{{\color{red} [Review-C: #1]}}
  \newcommand{\old}[1]{{\hl{#1}}}
\else

  \newcommand{\cjw}[1]{}
  \newcommand{\xuan}[1]{}
  \newcommand{\sean}[1]{}
  \newcommand{\todo}[1]{}
  \newcommand{\comment}[1]{}
  \newcommand{\old}[1]{}
\fi

\newcommand{\DesName}{\textit{Hercules}}

\begin{document}
\bstctlcite{IEEEexample:BSTcontrol}

\maketitle
\thispagestyle{plain}
\pagestyle{plain}


\begin{abstract}

Personalized recommendation is an important class of deep-learning applications that powers a large collection of internet services and consumes a considerable amount of datacenter resources.
As the scale of production-grade recommendation systems continues to grow, optimizing their serving performance and efficiency in a heterogeneous datacenter is important and can translate into infrastructure capacity saving.
In this paper, we propose {\DesName}, an optimized framework for personalized recommendation inference serving that targets diverse industry-representative models and cloud-scale heterogeneous systems.
{\DesName} performs a two-stage optimization procedure --- offline profiling and online serving.
The first stage searches the large under-explored task scheduling space with a gradient-based search algorithm achieving up to 9.0$\times$ latency-bounded throughput improvement on individual servers; it also identifies the optimal heterogeneous server architecture for each recommendation workload. 
The second stage performs heterogeneity-aware cluster provisioning to optimize resource mapping and allocation in response to fluctuating diurnal loads.
The proposed cluster scheduler in {\DesName} achieves 47.7\% cluster capacity saving and reduces the provisioned power by 23.7\% over a state-of-the-art greedy scheduler.

\end{abstract}

\section{Introduction}
\label{sec:intro}

The ability to predict user preference and provide meaningful experience is important for internet services, such as search engines, social networks, online retail and content streaming~\cite{GCP,youtube,Walmart_AI,Amazon_Personalize}.
Deep learning (DL)-based personalized recommendation models are the algorithmic engines that power these important services with high prediction accuracy and deliver quality user experiences~\cite{hazelwood2018applied}.
In order to capture fast-evolving data features, we have witnessed a variety of recommendation model innovations in the past few years~\cite{wnd,mtwnd,DLRM,din,dien}. 
Among other vision and language models often implemented with FC layers, CNN, or RNN, the deep learning recommendation systems at datacenters exhibit a number of unique workload characteristics and system requirements---model diversity, cloud-scale system heterogeneity, and time-varying load patterns---requiring an application-specific solution for execution efficiency~\cite{CrossStackRec}.

\begin{figure}[t!]
  \centering
  \includegraphics[width=\columnwidth]{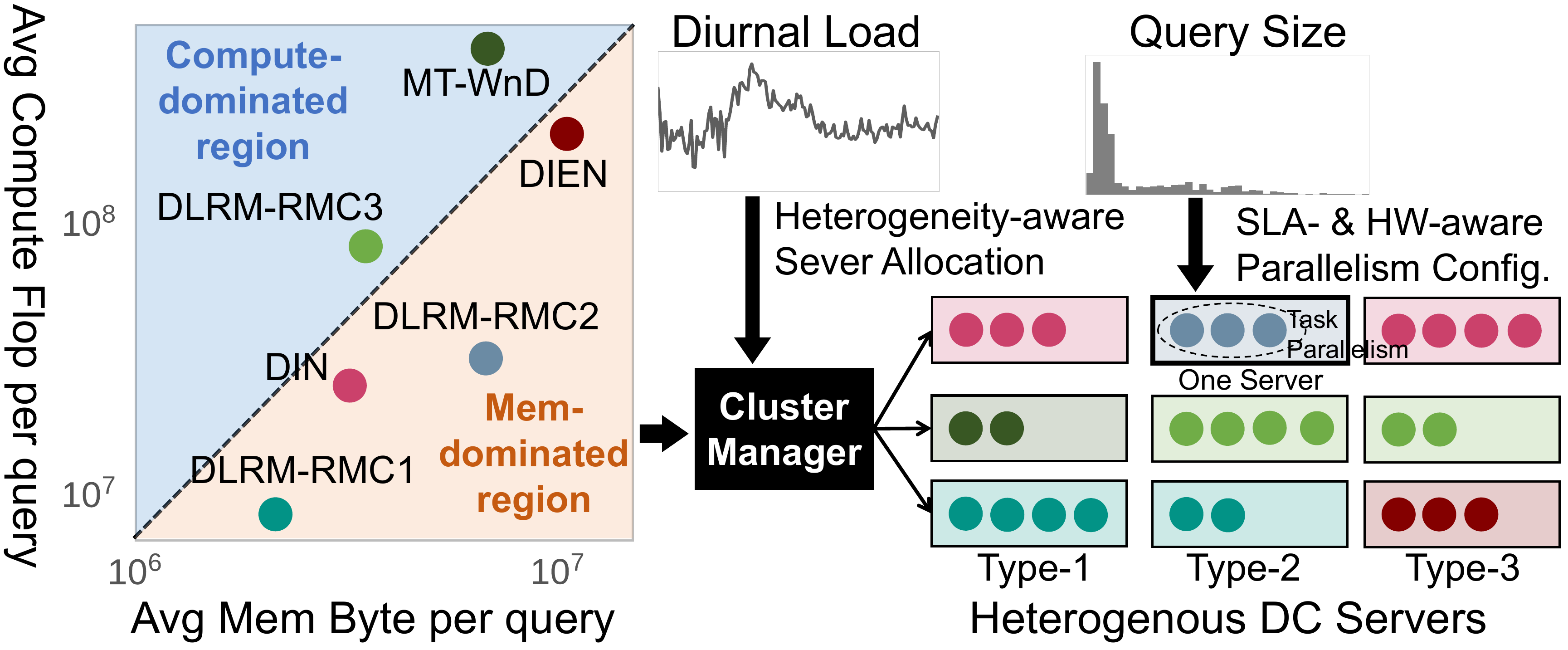}
  \vspace{-0.7cm}
  \caption{(Left) The compute and memory footprint of the industry-scale recommendation models varies significantly across use cases demanding distinct solutions;
  (Right) the cluster-level and server-level scheduling challenges that require schedulers to adapt system configurations and load variations.
  One box means one physical server.
  The different number of nodes in one physical server indicates the various parallelism configuration depending on model/server type to satisfy the SLA target.
  }
  \label{fig:motivation}
  \vspace{-0.7cm}
\end{figure}

\textbf{Model Diversity.} 
Depending on specific use cases, recommendation models are constructed differently for a wide variety of services. 
In 2019, Facebook had a few hundred recommendation models running concurrently in its datacenter fleet for inference serving~\cite{hpca-gupta-19}.
Moreover, the recommendation models can evolve rapidly to support new use cases and achieve higher prediction accuracy.
Such diverse algorithmic structures result in a varying spectrum of performance bottlenecks that fundamentally defy a ``one-size-fits-all'' solution. 
In Figure~\ref{fig:motivation}(left), the compute (y-axis) and memory intensity (x-axis) of state-of-the-art recommendation models can vary by one to two orders-of-magnitude. 
For instance, Google's MT-WnD~\cite{mtwnd} and Facebook's DLRM-RMC3~\cite{DLRM} are dominated by dense feature processing with wide FC layers, whereas Alibaba's DIN~\cite{din} and DIEN~\cite{dien} are dominated by attention units with FC and RNN layers, both consuming large computing resources.
In contrast, Facebook's DLRM-RMC1, RMC2~\cite{DLRM} are dominated by sparse feature processing that show lower compute intensity but higher memory bandwidth demand.

\textbf{Cloud-scale System Heterogeneity.}
A wide variety of system architectures can co-exist in modern datacenter fleets. 
Two primary reasons contribute to the increasing level of system heterogeneity.
First, system upgrades occur periodically, resulting in generations of servers with different micro-architectures~\cite{dc_heterogeneity}.
Second, domain-specific accelerators are increasingly deployed in datacenters to maximize execution efficiency~\cite{Accelerometer,Kelp}.
Deep learning (DL) accelerators, such as Nvidia GPUs, Google TPUs~\cite{tpu}, Facebook Kings Canyon~\cite{fb-intel,fb-accel}, Alibaba Hanguang~\cite{hanguang}, have been deployed to supplement traditional CPU-only industry-scale datacenters. 
In addition to compute-centric accelerators, near-memory processing (NMP) solutions~\cite{recnmp,tensordimm,Tensorcasting,FAFNIR,AxDIMM} 
have also been proposed to accelerate the memory-bounded operations which are specific to recommendation models.
This landscape of increasingly heterogeneous systems calls for heterogeneity-aware resource management infrastructures.

\textbf{Time-varying Load Patterns.}
Recommendation models are deployed across web services that exhibit timing-varying load patterns.
This dynamism manifests at both the server level and the cluster level as illustrated in Figure~\ref{fig:motivation}.
For individual servers, the query arrival pattern follows the Poisson distribution of query arrival rate and displays a distinct heavy-tail distribution of query sizes~\cite{DeepRecSys}.
Meanwhile, at the cluster level, we observe highly-fluctuating and synchronous diurnal loads for different recommendation services at the granularity of a day. This diurnal load pattern is typical for user-facing services~\cite{hazelwood2018applied}. 
The dynamically-changing conditions require the schedulers operating at different levels to quickly adapt and respond to the load variations.

\textit{Model diversity}, \textit{system heterogeneity}, and \textit{load variation} converge prominently in the design and optimization of recommendation inference serving systems---a significant consumer of cloud computing resources.
The combination of the aforementioned factors poses new challenges to at-scale recommendation inference execution and scheduling across datacenter fleets.
A task scheduler must intelligently partition/allocate the model execution to satisfy the strict tail-latency targets set by the Service Level Agreement (SLA).
Yet the optimal scheduling decision is highly model- and hardware-dependent, and requires an efficient search mechanism to fully explore the large scheduling space across the {model-, operator- and data-parallelism dimensions} for all SLA targets.
Existing task scheduler designs~\cite{DeepRecSys,Baymax,Prophet,lazybatching} lack the capability to traverse this full parallelism space.

To tackle these modern challenges, we propose a comprehensive optimization framework named {\DesName}---\underline{he}terogeneity-aware \underline{r}e\underline{c}ommendation \underline{u}sing \underline{l}atency- and \underline{e}nergy-conscious \underline{s}cheduler---tailor-designed for at-scale neural recommendation inferences.
To the best of our knowledge, it is the first work that \textit{jointly optimizes for model diversity, system heterogeneity, and dynamic load patterns} with a seamlessly integrated scheduler design at both the individual server and the cluster levels.
{\DesName} tackles cross-layer scheduling for latency-critical workloads at the datacenter scale---a problem not only central to efficient serving of heterogeneous recommendation inference but also generalizable to broader DL frameworks and at-scale inference scheduling.
Optimization in {\DesName} is performed in two stages.
In the offline profiling stage, {\DesName} exhaustively explores the parallelism space for task scheduling and derives an \textit{efficiency tuple} for all permutations of the workload/server type pairs.
In the online serving stage, {\DesName} performs heterogeneity-aware provisioning at the cluster level using the \textit{efficiency tuple} to solve the constrained optimization problem, minimizing the datacenter resource cost. Our work makes the following contributions:

\begin{itemize}
\setlength{\itemsep}{0pt}
\setlength{\parskip}{0pt}

\item  We formulate the wide design space into a constrained optimization problem and propose an efficient search method that effectively explores the parallelism space and identifies an optimal execution setting for heterogeneity-aware cluster scheduling.

\item We identify the under-explored parallelism space for task scheduling, achieving 1.03$\times$ to 9$\times$ latency-bounded throughput improvement over a state-of-the-art SLA-aware scheduler with DeepRecSys~\cite{DeepRecSys} for CPU and Baymax~\cite{Baymax} for accelerator on individual servers.

\item With real system measurement, {\DesName} can achieve up to 47.7\% cluster capacity and 23.7\% provisioned power saving over state-of-the-art greedy schedulers~\cite{paragon,Quasar}, leading to higher cluster-level resource efficiency.  

\end{itemize}

\section{Background and Related Works}
\label{sec:background}

\begin{figure*}[t!]
  \centering
  \includegraphics[width=\textwidth]{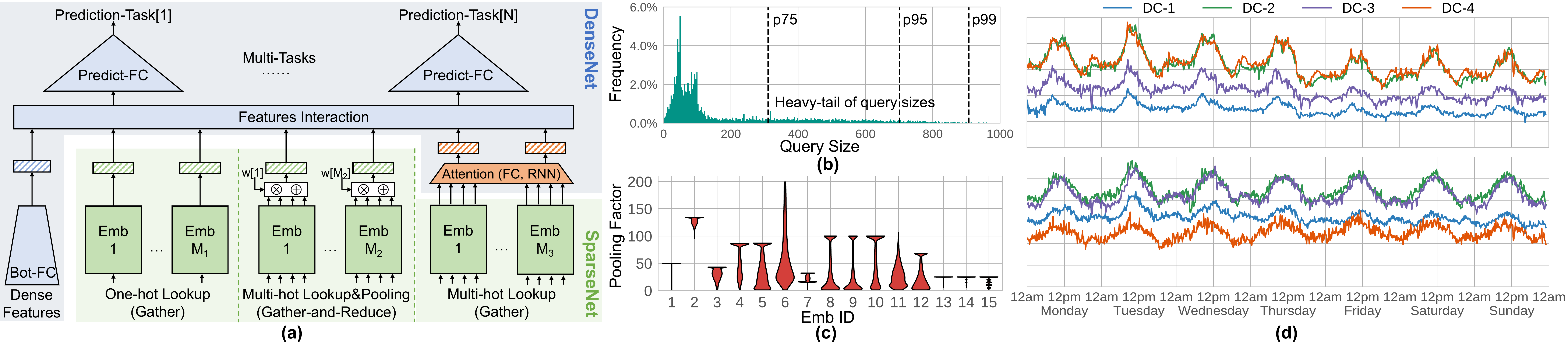}
  \vspace{-0.8cm}
  \caption{(a) General recommendation model architecture; 
  (b) Query sizes distribution of recommendation inference;
  (c) Pooling factor distribution of 15 embedding tables in 500 inference queries;
  (d) Diurnal load of two real industry services of four datacenters over one week period in March 2021.}
  \label{fig:rec_char}
  \vspace{-0.2cm}
\end{figure*}

\subsection{Industry-scale Recommendation Services}
\textbf{Diverse Recommendation Models.}
DL-based recommendation models are widely used to improve the quality of user experiences for internet services. 
In 2019, recommendation inferences consumed roughly 80\% of the total machine learning cycles at Facebook~\cite{hpca-gupta-19}, with similar trends observed by Google, Alibaba, and Amazon~\cite{youtube,alibabaRec,Amazon_Personalize}.
Figure~\ref{fig:rec_char}(a) shows a typical recommendation model consisting of a SparseNet with memory-intensive sparse operations on embeddings and a DenseNet
with compute-intensive operations. 
Our study uses six recommendation models~\cite{DLRM,mtwnd,din,dien} summarized in Table~\ref{tab:rm_architecture} representing the predominant model architectures employed by popular services in industries.

\textbf{Dynamically-varying Working Set Sizes.}
The query size of recommendation inference represents the number of items to be ranked for a user~\cite{DeepRecSys} and is heavily dependent on the user's interaction with the web service.
Figure~\ref{fig:rec_char}(b) shows the query size histogram from a production recommender system, typically varying between 10 and 1000.
Moreover, the number of the embedding entries in one embedding lookup and pooling operation exhibits large variance and high dependence on the sparse features.
Figure~\ref{fig:rec_char}(c) provides an example distribution of pooling factors across 15 embedding tables in 500 production inference queries.
At-scale recommendation services face unique task scheduling challenges that arise from the dynamic variation and a heavy tail of the working set size.

\begin{table*}[t!]
\centering

\scriptsize
\caption{State-of-the-art production-scale recommendation models configurations}
\vspace{-0.2cm}

\label{tab:rm_architecture}

\begin{tabular}{|c|c|c|c|c|c|c|c|c|c|}
\hline
\multirow{3}{*}{Models} &
  \multirow{3}{*}{Services} &
  \multicolumn{5}{c|}{SparseNet} &
  \multirow{3}{*}{Attention} &
  \multicolumn{2}{c|}{DenseNet} \\ \cline{3-7} \cline{9-10} 
 &
   &
  \multirow{2}{*}{\# of Embs} &
  \multicolumn{2}{c|}{Emb Size} &
  \multirow{2}{*}{Lookups} &
  \multirow{2}{*}{Pooling} &
   &
  \multirow{2}{*}{Bottom-FC} &
  \multirow{2}{*}{Predict-FC} \\ \cline{4-5}
          &              &                    & Prod        & Small     &               &     &     &             &                         \\ \hline
DLRM-RMC1 & Social Media & $\sim$10 $\times$  & 1M - 5M     & 1M        & 20 - 160      & Yes & -   & 256-128-32  & 256-64-1                \\ \hline
DLRM-RMC2 & Social Media & $\sim$100 $\times$ & 1M - 5M     & 1M        & 20 - 160      & Yes & -   & 256-128-32  & 512-128-1               \\ \hline
DLRM-RMC3 & Social Media & $\sim$10 $\times$  & 10M - 20M   & 1M        & 20 - 50       & Yes & -   & 2560-512-32 & 512-128-1               \\ \hline
MT-WnD    & Video        & 26                 & 3 - 40M     & 1M    & 1             & No  & -   & -           & N$\times$(1024-512-256) \\ \hline
DIN       & E-commerce   & 3                  & 0.1M - 600M & 0.1M - 1M & 1, 100 - 1000 & No  & FC  & -           & 200-80-2                \\ \hline
DIEN      & E-commerce   & 3                  & 0.1M - 600M & 0.1M - 1M & 1, 100 - 1000 & No  & GRU & -           & 200-80-2                \\ \hline
\end{tabular}
\vspace{-0.4cm}
\end{table*}

\textbf{Diurnal Load of Recommendation Services.}
To serve billions of users around the world, datacenters maintain a large fleet of servers to handle the peak load at any given time.
The arrival queries per second of recommendation services follow a diurnal pattern~\cite{hazelwood2018applied}.
Figure~\ref{fig:rec_char}(d) shows the real temporal loads of two recommendation services arriving at industry datacenters and their synchronous diurnal pattern.  
For both services, all four datacenters reach peak/valley loads around similar times.
Moreover, the two distinct services also display synchronous diurnal patterns. 
This synchronous diurnal pattern across datacenters and services leads to $>$50\% fluctuation from the aggregated loads between peak and off-peak times.
The unbalanced peaks with high amplitudes can result in imprudent over-provisioning of resources to satisfy the peak load yet poor utilization of the allocated resources off peak.

\subsection{Task Scheduling on Individual Servers}

\begin{figure}[t!]
  \centering
  \includegraphics[width=\columnwidth]{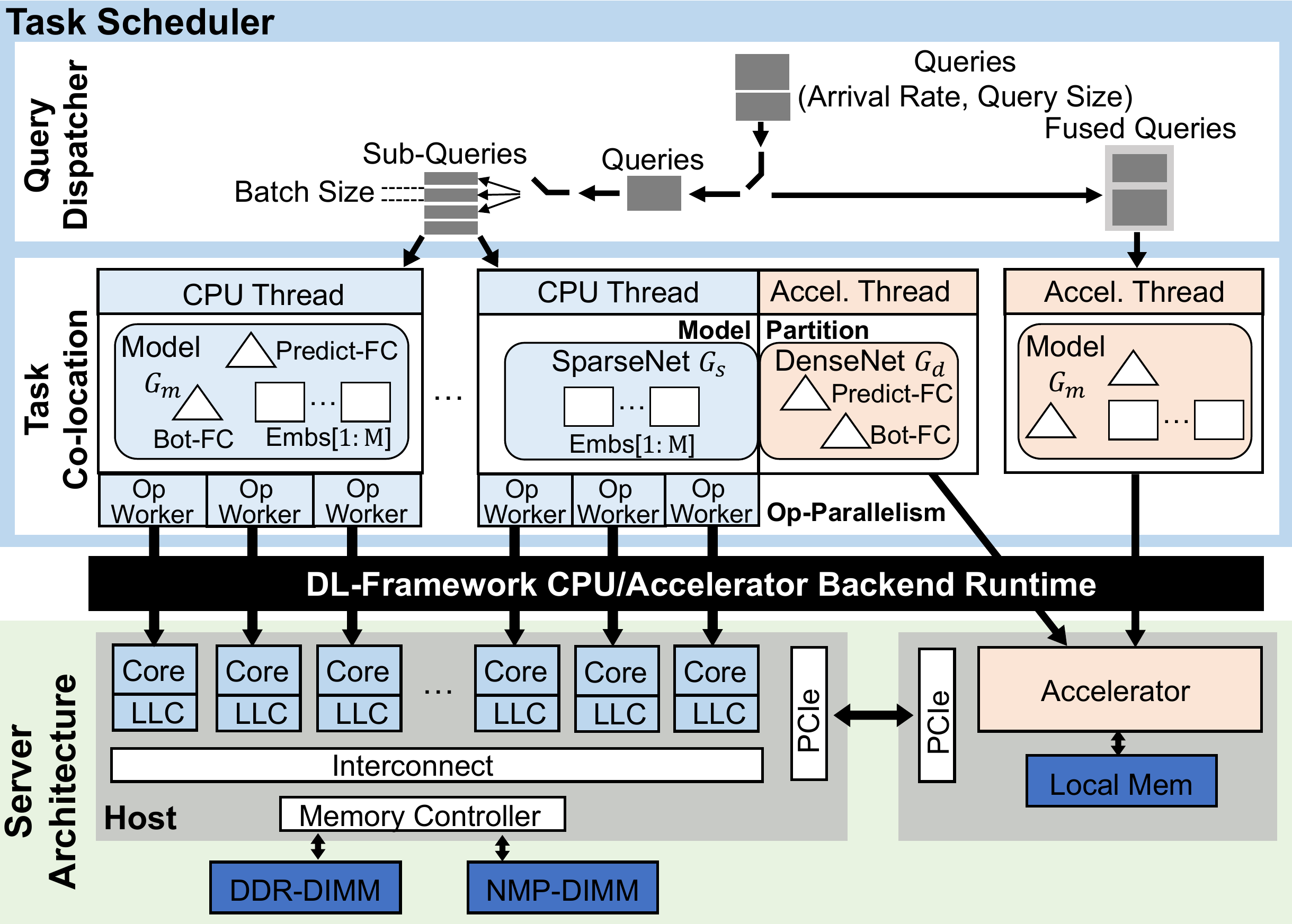}
  \vspace{-0.6cm}
  \caption{Overall system stack in data-center scale recommendation inference. 
  The task scheduler manages co-location of recommendation models across multi-core CPUs and accelerators by co-designing solutions based on characteristics of the workload, query dispatcher, and underlying server architecture.  }
  \label{fig:system_stack}
  \vspace{-0.4cm}
\end{figure}

\textbf{System Stack.}
To serve recommendation inferences, the system stack consists of three abstract layers---task scheduler,
deep learning (DL-)framework, 
and server architecture.
As shown in Figure~\ref{fig:system_stack}, the task scheduler 
manages the co-located inference threads on server-grade CPUs and accelerators.
The DL-framework acts as an interface between the task scheduler and the underlying hardware devices.
Typical DL-based models can be represented as computation graphs in DL frameworks such as PyTorch~\cite{pytorch} and Caffe2~\cite{caffe2}.
The model can then be launched as a whole end-to-end computation graph $G_m$ by one inference thread, referred to as \textit{model-based scheduling} in this paper. It can also be partitioned into multiple sub-graphs, for example, a SparseNet $G_s$ and a DenseNet $G_d$, and launched by two inference threads in a pipelined fashion.
Each inference thread activates one graph executor instantiated by the DL-framework.
Based on the operator dependency defined by the computation graph, the graph executor launches the operators in order, and parallel operator workers can be assigned to one graph executor to launch independent operators in parallel.

\textbf{Parallelism Space.}
Supported by the system stack, the task scheduler exploits a large parallelism space, including \textit{model-, operator (op-), and data-parallelism}.
\textit{Model-parallelism} represents the number of parallel inference threads,  either the end-to-end model $G_m$ inference threads in \textit{model-based scheduling} or the sub-graphs (e.g., $G_s$ and $G_d$) inference threads with model partition and pipelining.
On multi-core CPUs, the inference threads are statically assigned to the physical cores without hyperthreading.
On accelerators, the inference threads are launched concurrently on a single accelerator, as \textit{model co-location}.
For example, Nvidia MPS (Multi-Process Service) scheduling~\cite{nvidia-mps} enables concurrent sharing of a GPU among multiple tasks.
\textit{Op-parallelism} represents parallel operator workers assigned to one inference thread and is only feasible on multi-core CPUs, where one physical core is allocated for one operator worker.
\textit{Data-parallelism} represents the inference batch size configured to serve the incoming queries in parallel.
On multi-core CPUs, each large inference query is split into multiple sub-queries and the query dispatcher distributes sub-queries to the parallel inference threads.
On accelerators, the inference queries are fused into one large batch to launch in parallel, referred to as \textit{query fusion}.

\textbf{SLA-aware Scheduling.}
For latency-critical recommendation workloads, the three dimensions of parallelism result in a large exploration space for the task scheduler to select the optimal parallelism configuration and maximize the throughput while satisfying the strict SLA latency target.
Since the optimal parallelism configuration is highly model- and hardware-dependent, a comprehensive and efficient exploration of parallelism space is needed for every permutation of the workload/server architecture pair.

Previous works~\cite{OptCNN,SOAP,PipeDream} explored model partition and pipelining for DNN training on GPU clusters.
These methods, however, cannot be directly applied to inference serving, as the performance metric of training is different from inference which needs to meet SLA latency targets. 
Moreover, model partitioning of DNN training across GPUs leverages large batch sizes to achieve high resource utilization.
It takes advantage of the pre-determined nature of inputs in training to form large input batches, but such opportunity is unavailable in inferences with dynamic arrival patterns where large batches can lead to long queuing delay.

On the other hand, SLA-aware scheduling mechanisms were proposed to exploit batching ({\em i.e.,} \textit{data-parallelism}) and model co-location ({\em i.e.,} \textit{model-parallelism}) on multi-core CPUs~\cite{DeepRecSys} and accelerators~\cite{Baymax,Prophet,DeepRecSys,lazybatching} to provide tail-latency guarantees.
Baymax~\cite{Baymax} performs QoS-aware management of the co-located tasks on accelerators by kernel reordering. 
DeepRecSys~\cite{DeepRecSys} is the most relevant method studying DL-based recommendation workloads.
LazyBatching~\cite{lazybatching} proposes fine-grain node-level batching, which is orthogonal to this work.
As our characterization will demonstrate in Section~\ref{sec:characterize}, the coverage of the parallelism space explored in these prior works remains limited, leaving large room for performance optimization.

\subsection{Cluster Scheduling at Datacenter Scale}

To manage a heterogeneous cluster, the cluster scheduler first classifies workloads with respect to heterogeneity based on the workload's performance on distinct server configurations.
It then determines and allocates the appropriate number of available best-matching servers to the incoming loads.

\textbf{Workload Classification.}
To predict and rank workload performance on different server architectures, 
previous works rely on detailed offline profiling~\cite{Bubble-Up} or online analytical prediction~\cite{paragon}.
Although low-level metrics such as CPU utilization, instructions per second (IPS), and
interference of shared resources can be used for workload classification~\cite{paragon,Quasar,Bubble-Up,Bubble-Flux}, these methods do not translate universally to server architectures that employ abundant heterogeneous accelerators.
Instead, high-level metrics, such as the achievable latency-bounded throughput (QPS) for latency-critical workloads, are more representative and are able to capture performance across different server configurations.

\textbf{Scheduling Policy.}
Once the workloads are accurately classified and ranked, prior work applies a greedy scheduler~\cite{paragon,Quasar} which always picks the available best-performing servers.
Then, the number of the servers to be activated is dynamically determined by the transient diurnal loads. 
During peak traffic, if the number of highest-ranked servers is not sufficient, the scheduler will start allocating lower-ranked servers for enough capacity to serve the incoming loads.
Similarly, during off-peak time, the scheduler first releases the amount of servers based on the descending order of servers' rankings. 
This greedy approach, however, lacks well-formulated global objectives, and therefore is not guaranteed to achieve cost minimization, as revealed by our investigation in Section~\ref{sec:char_cluster}.

\section{Characterization}
\label{sec:characterize}
We perform an in-depth characterization of task scheduling and cluster scheduling on recommendation inference workloads.
The analysis uncovers several previously untapped performance optimization opportunities, and quantifies the limitation of existing scheduling methods.
We construct an SLA-aware task scheduler---DeepRecSys~\cite{DeepRecSys} to exploit the \textit{data-parallelism} across general-purpose CPUs and Baymax~\cite{Baymax} to explore the \textit{model co-location} on GPU-based accelerators---as the baseline.
We use the hill-climbing algorithm in~\cite{DeepRecSys} searching for the optimal scheduling configuration on both CPUs and GPUs that maximizes latency-bounded throughput.
We observe that the potential benefits come from the extensive exploration of the additional scheduling design spaces among CPUs, accelerators and heterogeneous-aware cluster manager.
We also find that the thorough exploration of the additional scheduling space provides up to 1.35$\times$ (CPUs) and 3.58$\times$ (GPUs) improvement on latency-bounded throughput and 1.33$\times$ (CPUs) and 2.11$\times$ (GPUs) savings on energy efficiency in individual servers.
Moreover, our scheduler can save the global cluster capacity by up to 41.6\% and the provisioned power by up to 11.4\%.
All system configurations used in this characterization are listed in Table~\ref{tab:sys_config} (Section~\ref{sec:method}).

\subsection{Host-side (CPU) Task Scheduling}
\label{sec:char_host_space}
\begin{figure}[t!]
  \centering
   \includegraphics[width=\columnwidth]{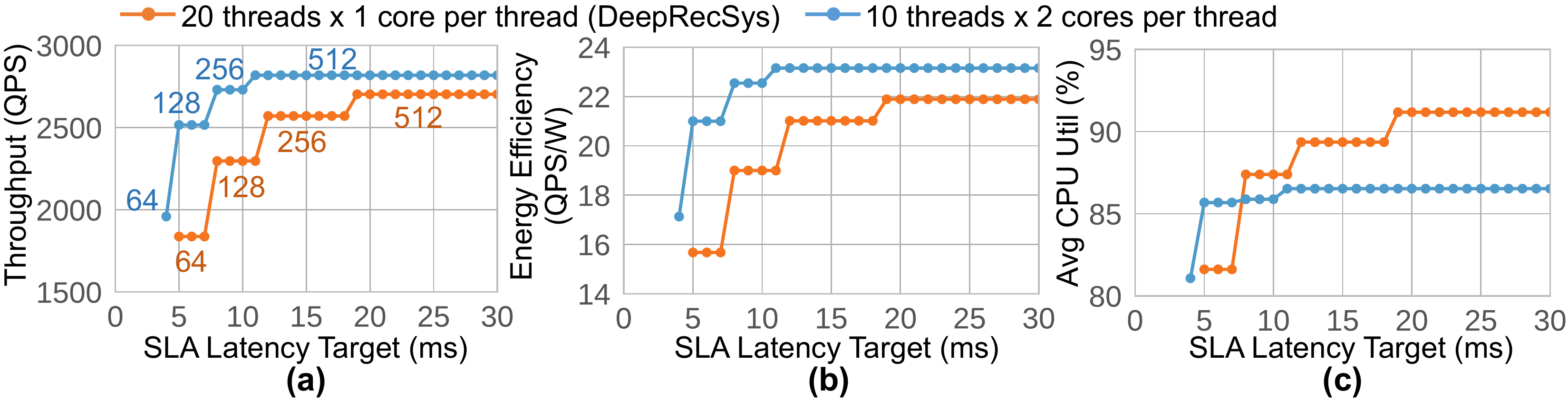}
  \vspace{-0.8cm}
  \caption{
  Host-side (a) latency-bounded throughput (QPS), (b) energy efficiency (QPS-per-Watt) and (c) average CPU utilization of DLRM-RMC1.}
  \label{fig:cpu_latency_bounded_data}
  \vspace{-0.2cm}
\end{figure}

First, we explore the benefits of task scheduling on multi-core CPUs by exploiting \textit{model- and op-parallelism}.
To explore the parallelism space, we compare two configurations running on a CPU-T2 server type Intel Xeon processor with 20 physical cores (see Table~\ref{tab:sys_config}).
The first configuration represents DeepRecSys~\cite{DeepRecSys} with fixed 20 inference threads and 1 core per thread (20$\times$1) while the second one uses 10 inference threads $\times$ 2 cores per thread (10$\times$2).
As shown in Figure~{\ref{fig:cpu_latency_bounded_data}}, DeepRecSys can satisfy a tight SLA latency target with smaller batch sizes while the 10$\times$2 configuration can leverage the \textit{op-parallelism} and reduce interference with half co-located inference threads, improving the latency-bounded throughput and energy efficiency by up to 35\% and 33\%, respectively.
Furthermore, while low-level metrics~\cite{Bubble-Up,paragon,Heracles} such as CPU utilization are commonly used to measure compute efficiency, Figure~\ref{fig:cpu_latency_bounded_data}(c) shows that CPU utilization is not directly correlated to the DL inference performance.
This suggests that high-level metrics ({\em e.g.,} latency-bounded throughput and power consumption) are more indicative than low-level metrics ({\em e.g.,} CPU utilization) for workload classification in later cluster scheduling steps.

\begin{figure}[t!]
  \centering
   \includegraphics[width=\columnwidth]{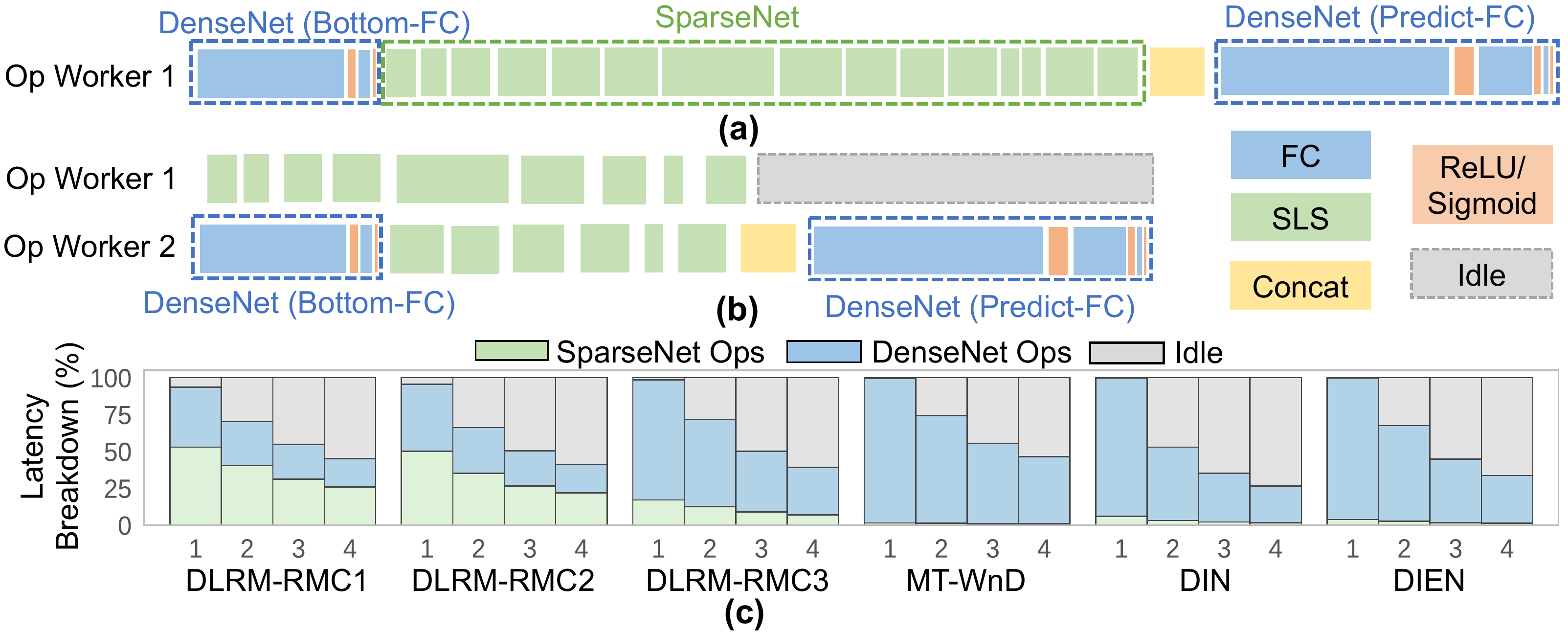}
  \vspace{-0.8cm}
  \caption{
  Host-side inference of DLRM-RMC1 with 
  (a) single operator worker per thread and 
  (b) two parallel operator workers per thread. 
  (c) Latency breakdown of the 6 models (batch size = 256) with 1, 2, 3 and 4 parallel operator workers per thread}
  \label{fig:cpu_char_breakdown}
  \vspace{-0.6cm}
\end{figure}

In addition to exploiting \textit{model-parallelism}, we find that there is significant room for performance improvement by balancing \textit{op-parallelism}.
While the 10$\times$2 configuration outperforms the 20$\times$1 configuration, it suffers from low CPU utilization as shown in Figure~\ref{fig:cpu_latency_bounded_data}(c).
This is a result of overheads from host-side inference operators' dependency and imbalanced workload distribution among the parallel operator workers.
Figure~\ref{fig:cpu_char_breakdown}(a) and (b) show the operator latencies of a single operator worker versus two parallel operator workers for DLRM-RMC1.
As can be seen, the inference thread with two parallel operator workers results in long idle time.
In Figure~\ref{fig:cpu_char_breakdown}(a), only one operator worker is allocated for the inference thread to process operators sequentially. 
For two workers in Figure~\ref{fig:cpu_char_breakdown}(b), 
due to operators' dependency, the Predict-FC cannot start before the Bottom-FC and the SparseNet finish, 
leaving one operator worker idle.
Figure~\ref{fig:cpu_char_breakdown}(c) shows that as the number of parallel workers increases, the idle cycles increase linearly for all six models.
The idle cycles range from 25\% to 74\% with 2 to 4 parallel operator workers.

\textit{
Insights:
under-exploration of parallelism dimension and workload imbalance in model-based scheduling leave significant rooms for further improvements in latency-bounded throughput (QPS) and energy efficiency (QPS-per-Watt).}

\subsection{Accelerator-side Task Scheduling}
\label{sec:char_accel_space}

\begin{figure}[t!]
  \centering
   \includegraphics[width=\columnwidth]{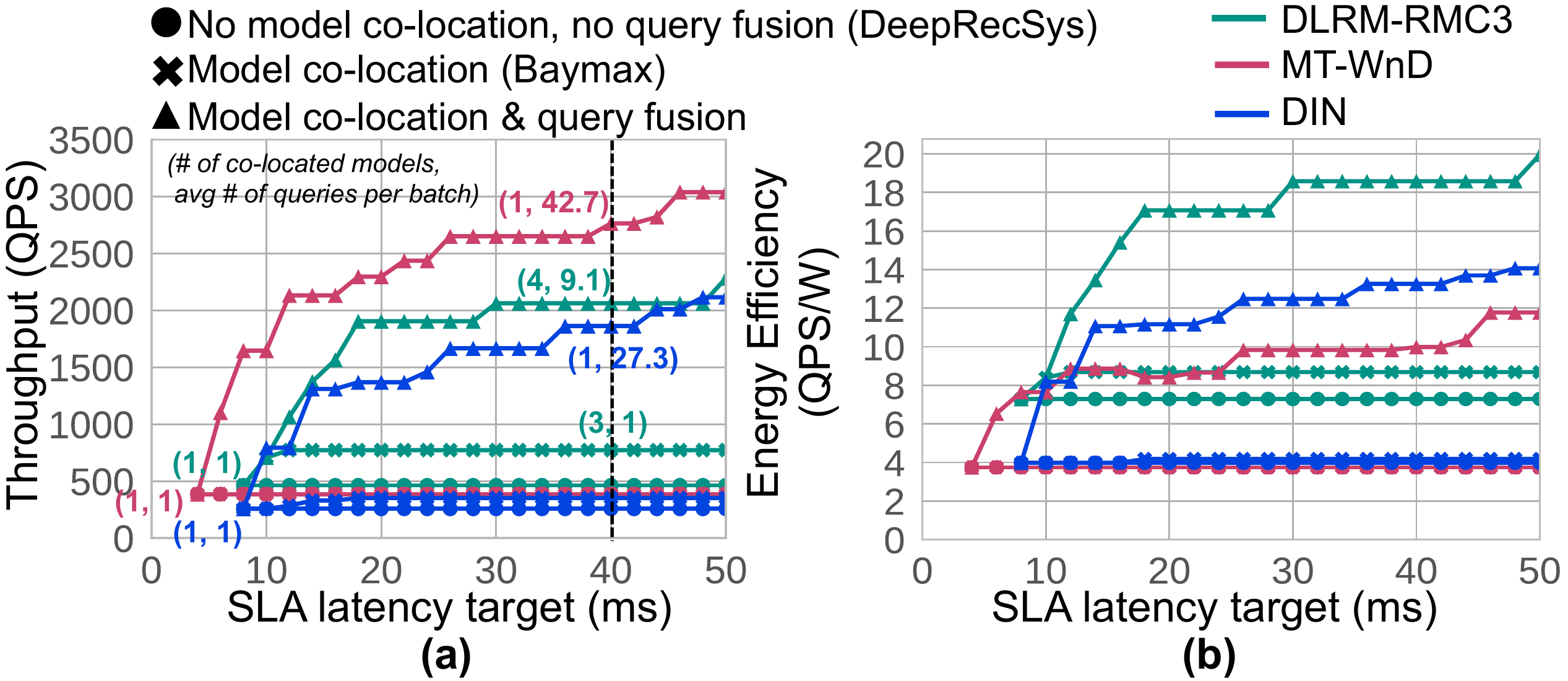}
  \vspace{-0.8cm}
  \caption{Latency-bounded (a) throughput and (b) energy efficiency (b) of three recommendations models (DLRM-RMC3, MT-WnD, DIN) with three accelerator-side task scheduling policies: no model co-location and no query fusion (DeepRecSys), only model co-location (Baymax), and both model co-location and query fusion. 
  Both model co-location and query-fusion provide significant performance and efficiency benefits over the baseline approaches.}
  \label{fig:gpu_latency_bounded_data}
  \vspace{-0.2cm}
\end{figure}

\begin{figure}[t!]
  \centering
   \includegraphics[width=\columnwidth]{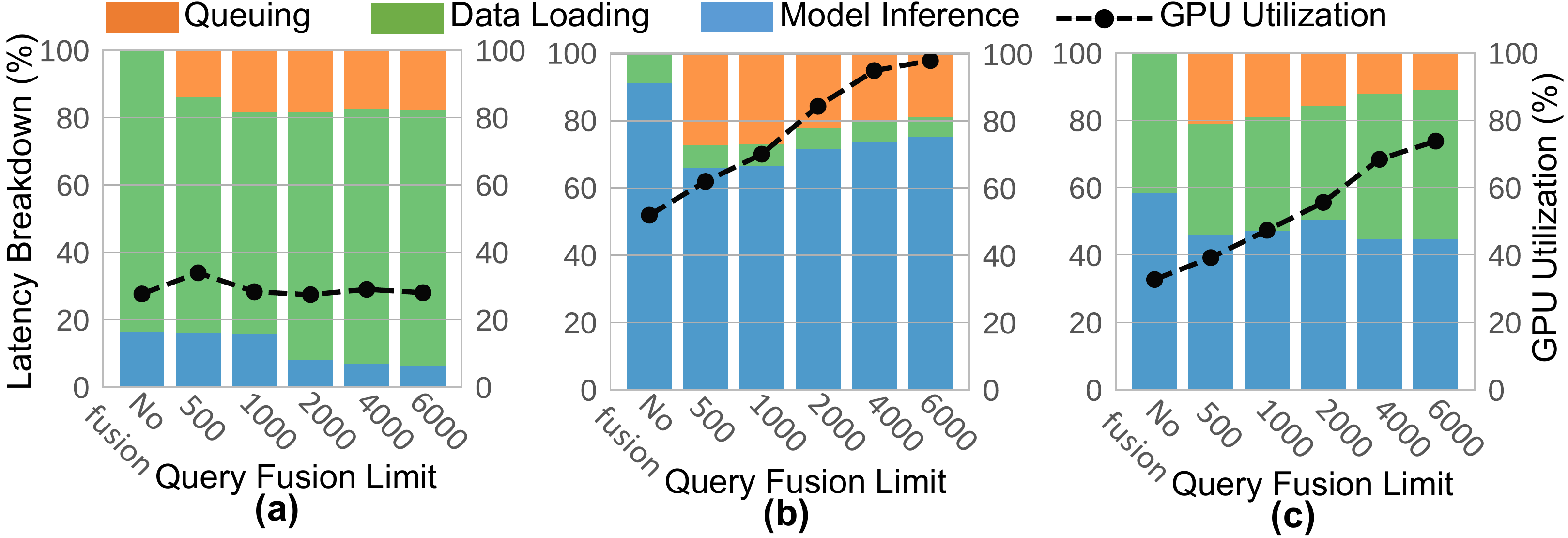}
  \vspace{-0.8cm}
  \caption{Breakdown of latency between queuing delay, data-loading, model inference, also GPU utilization for (a) DLRM-RMC3, (b) MT-WnD, (c) DIN.}
  \label{fig:gpu_latency_breakdown}
  \vspace{-0.6cm}
\end{figure}

Next, we investigate the scheduling opportunities on the accelerator side. 
Due to accelerators' limited memory capacity (16GB in Nvidia V100), the \textit{model-based scheduling}~\cite{DeepRecSys,Baymax} method does not scale to large recommendation models; therefore, only the smaller versions of the models in Table~\ref{tab:rm_architecture} are used for characterization.
Here, \textit{model-parallelism} manifests as \textit{model co-location} by increasing the number of co-located inference threads on a single accelerator, and \textit{data-parallelism} manifests as \textit{query fusion} by merging multiple queries into one batch processed in parallel.
As shown in Figure~\ref{fig:gpu_latency_bounded_data}, \old{three parallelism configuration used in our evaluations are 
DeepRecSys~\cite{DeepRecSys} with no model co-location and no query fusion,
Baymax~\cite{Baymax} with model co-location, and a third approach we constructed that combines both model co-location and query fusion. }
we evaluate three parallelism configurations: (1) DeepRecSys~\cite{DeepRecSys} with no model co-location and no query fusion, (2) Baymax~\cite{Baymax} with model co-location only, (3) an approach we contrived combining both model co-location and query fusion.
Using DLRM-RMC3, MT-WnD, and DIN models, we observe that \textit{model co-location} in Baymax improves the latency-bounded throughput by up to 1.66$\times$/1.03$\times$/1.36$\times$ and 
energy efficiency by up to 1.19$\times$/1.02$\times$/1.06$\times$ over DeepRecSys.
The joint exploration of \textit{model co-location} and \textit{query fusion} reveals additional performance improvement.
We observe a latency-bounded throughput improvement of up to 2.95$\times$/7.87$\times$/6.0$\times$ and 
energy efficiency improvement of up to 2.29$\times$/3.14$\times$/3.36$\times$
over Baymax.

Distinct from the performance interference observed on multi-core CPUs, the performance of hardware accelerators is often degraded by the queuing delay from waiting for multiple queries to form one batch and PCIe bandwidth contention from data loading between the host and the accelerators. 
More specifically, Figure~\ref{fig:gpu_latency_breakdown} shows the latency breakdown of the queuing delay and the other two pipelined stages (data loading and model inference) for three models with one single inference thread on one Nvidia V100.
For DLRM-RMC3, 65--83\% of the end-to-end latency is contributed by data loading to transfer a large number of sparse indices for multi-hot embedding operations.
This results in long GPU idle times and low utilization at $\sim$25\%.
MT-WnD and DIN are better at keeping the GPU busy as the one-hot embedding lookup in MT-WnD incurs lower overhead in data loading and the Attention-Net in DIN is compute-intensive, mitigating the data loading overhead.

\textit{Insights:
Accelerators like GPUs also benefit from concurrent exploration of model co-location and query fusion, achieving 7.87$\times$ throughput and 3.36$\times$ energy efficiency improvement over a state-of-the-art scheduler, yet they cannot accommodate large models due to limited memory capacity.}

\subsection{Heterogeneity-aware Cluster Scheduling}
\label{sec:char_cluster}

\begin{figure}[t!]
  \centering
   \includegraphics[width=\columnwidth]{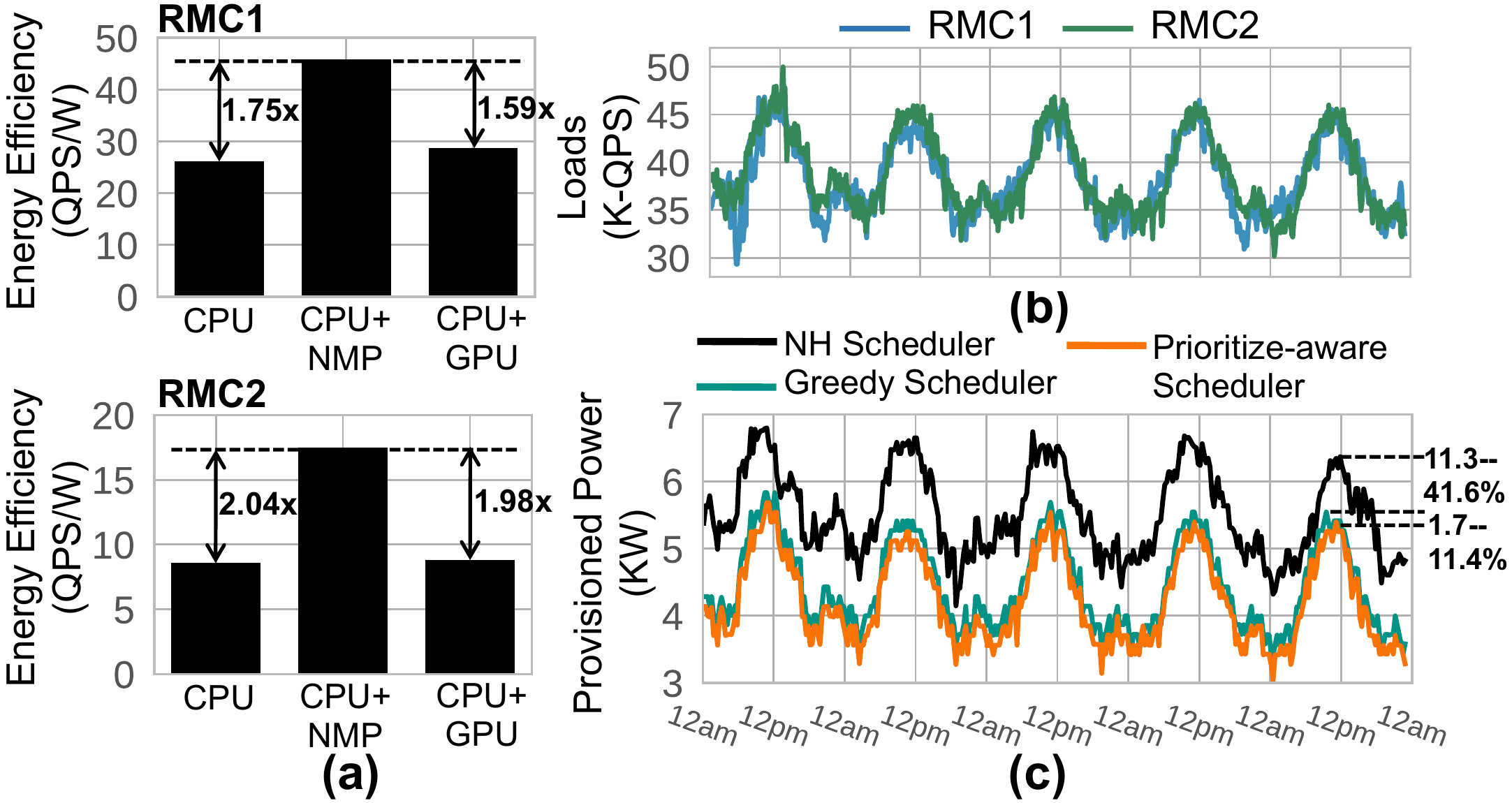}
  \vspace{-0.6cm}
  \caption{
  (a) The latency-bounded energy efficiency of DLRM-RMC1 and RMC2 on the three server types.
  (b) Loads of DLRM-RMC1 and RMC2;
  (c) Provisioned power budget of heterogeneity-oblivious (NH), greedy scheduler from~\cite{paragon,Quasar}, and prioritize-aware scheduler.
  }
  \label{fig:cluster_efficiency}
  \vspace{-0.4cm}
\end{figure}

Resources in datacenters are often over-provisioned to satisfy peak load, and one important goal of improving cluster efficiency is to reduce the amount of over-provisioned power.
As introduced in Section~\ref{sec:background}, prior works~\cite{paragon,Quasar} proposed heterogeneity-aware provisioning to dynamically activate the needed number of best-matching servers for current loads.
To compare and contrast different cluster scheduling schemes, we consider a heterogeneous cluster consisting of three types of servers, CPU-only, CPU+NMP, and CPU+GPU (server type $T_{2}$, $T_{3}$ and $T_{7}$ in Table~\ref{tab:sys_config}) and assume their respective availability is 70, 15 and 5.
The cluster is set up to serve workloads of DLRM-RMC1 and DLRM-RMC2.

Our characterization looks at the performance of two existing cluster scheduling schemes from~\cite{paragon,Quasar}---the heterogeneity-oblivious (NH) scheduler and the heterogeneity-aware greedy scheduler.
First, workload classification is performed to rank the available server types according to their latency-bounded throughput/energy efficiency. 
Given the SLA target of 20ms and 50ms for RMC1 and RMC2, the server candidates are ranked as CPU+NMP$>$CPU+GPU$>$CPU for both workloads based on energy efficiency (QPS/W) as shown in Figure~\ref{fig:cluster_efficiency}(a).
During online serving, the workloads follow the diurnal pattern as shown in Figure~\ref{fig:cluster_efficiency}(b) and each has a peak load of 50K.
The NH scheduler does not consider system heterogeneity and randomly assigns the available servers for the incoming loads, whereas the greedy scheduler assigns the highest-ranked available servers, achieving provisioned power saving by 41.6\% at peak and 21.5\% on average.
However, the greedy policy randomly divides the highest-ranked CPU+NMP servers between RMC1 and RMC2 and thus misses additional optimization opportunity.
As shown in Figure~\ref{fig:cluster_efficiency}(a), CPU+NMP achieves higher energy efficiency improvement on RMC2 over RMC1.
More provisioned power can be saved by prioritizing the allocation of CPU+NMP servers to RMC2.
This observation leads us to construct the priority-aware scheduler which can achieve additional power saving of 11.4\% at peak and 4.2\% on average over a greedy scheduler, as illustrated in Figure~\ref{fig:cluster_efficiency}(c).

\textit{Insights:
Despite being heterogeneity-aware, the state-of-the-art greedy scheduler fails to correctly arbitrate the server allocation when multiple workloads compete for the same server types, resulting in sub-optimal solution.
}

\section{Hercules Design}
\label{sec:design}

\begin{figure*}[t!]
  \centering
  \includegraphics[width=\textwidth]{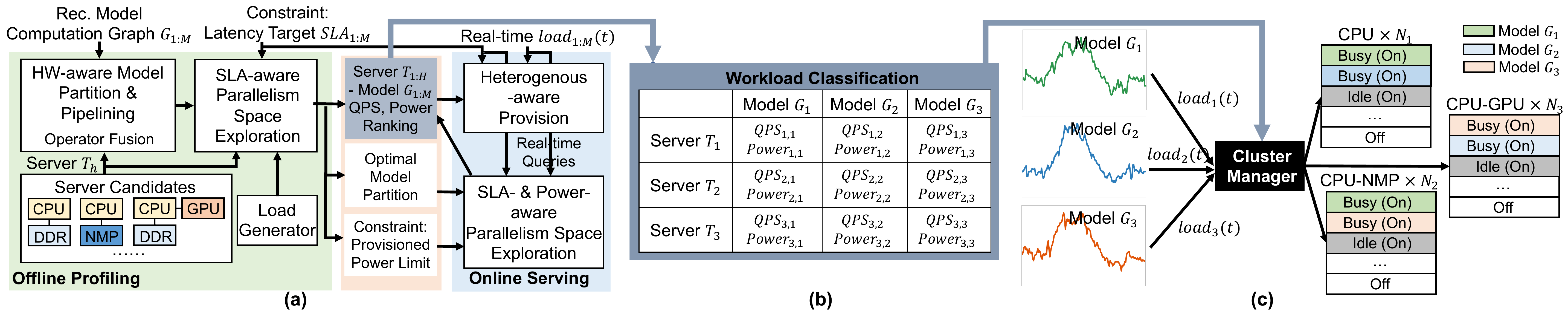}
  \vspace{-0.6cm}
  \caption{(a) {\DesName} two-stage optimization flow;
  (b) Workload classification table;
  (c) Heterogeneous-aware provision.}
  \label{fig:design_flow}
  \vspace{-0.4cm}
\end{figure*}

\subsection{Overview of {\DesName} Optimization Framework}
The key insights unraveled by our characterization guided the design of the {\DesName} optimization framework.
{\DesName} consists of two main stages: offline profiling and online serving.
During offline profiling, we aim to maximize the efficiency of recommendation workload execution on individual servers, and record an \textit{efficiency tuple} for all workload/server type pairs; this workload classification metric is used subsequently to guide the cluster scheduler during online serving. 
As illustrated in Figure~\ref{fig:design_flow}(a), {\DesName} takes the recommendation model in the form of a computation graph $G_m$ and its corresponding SLA latency target $SLA_m$.
The offline profiling is performed by evaluating every server candidate $T_h$ for model $G_m$.
First, for every $G_m$ and $T_h$ pair, a hardware-(HW-)aware model partition is performed to satisfy the memory capacity constraint in hardware for large recommendation models;
then the SLA-aware task scheduling exploration is performed to achieve the maximal latency-bounded throughput while satisfying the SLA latency target.
As shown in Figure~\ref{fig:design_flow}(b), the latency-bounded throughput and measured peak power ($QPS_{m,h}$, $Power_{m,h}$) are recorded as the \textit{efficiency tuple} to quantitatively classify the available server architectures for each workload.
Also, the offline measured peak power $Power_{m,h}$ is used as the provisioned power budget for the online allocated servers.
During online serving, initial setup is first performed by running the SLA- and power-aware task scheduling exploration to ensure accurate profiling with the real-time queries.
Two constraints must be satisfied here---the SLA latency target ($SLA_m$) is met and the power consumption is within the provisioned power budget $Power_{m,h}$.
The \textit{efficiency tuple} is also updated in real-time to reflect the measured performance $QPS_{m,h}$ with real-time query loads.
With the quantitative workload classification, we formulate the cluster provision as a constrained optimization problem for a global resources cost minimization objective.
The right amount of best-matching servers are dynamically allocated satisfying the incoming diurnal loads.

\subsection{Gradient-guided Task Scheduling Exploration}
Recommendation inference on an individual server needs to satisfy three constraints: hardware resource, SLA latency, and provisioned power budget. 
Therefore, {\DesName} performs HW-aware model partition to comply with the available memory capacity and parallelism space exploration to meet the SLA latency and power constraint.

\begin{figure}[t!]
  \centering
  \includegraphics[width=\columnwidth]{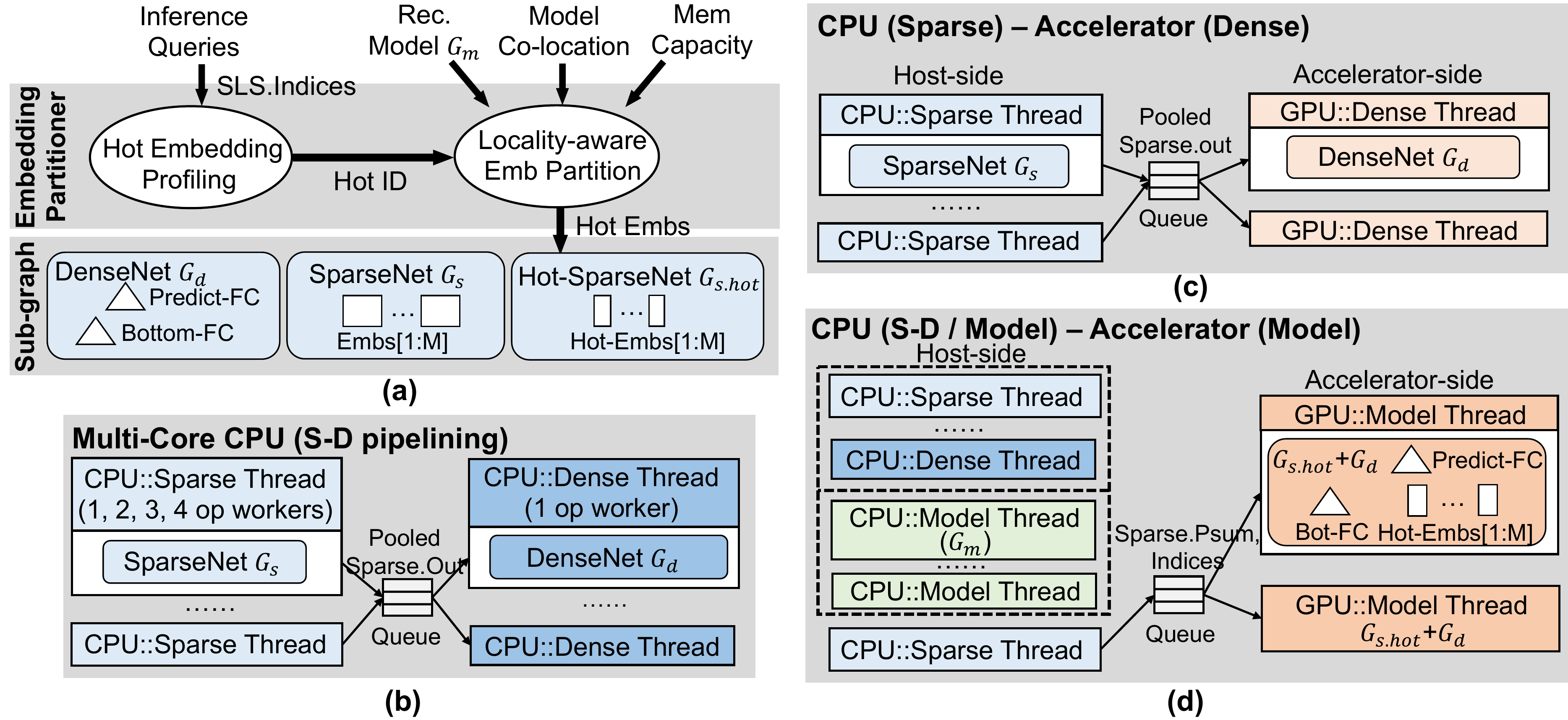}
  \vspace{-0.7cm}
  \caption{(a) Graph partition; 
  (b) \textit{S-D pipeline scheduling} on CPU platform; 
  (c) \textit{S-D pipeline scheduling} on CPU-Accelerator platform;
  (d) \textit{S-D pipeline scheduling} or \textit{model-based scheduling} on the host side and \textit{model-based scheduling} on the Accelerator side.
  }
  \label{fig:graph_partition}
  \vspace{-0.4cm}
\end{figure}

\textbf{HW-aware Model Partition and Pipelining.}
Given the sheer size of production-scale recommendation models, {\em model partition} is performed before they can feasibly be executed on the accelerators of limited memory capacity.
First, we found that more than  95\% of the model memory footprint comes from the embedding tables in SparseNet $G_s$. 
DenseNet $G_d$ only consumes a few MBs and can be easily held on the accelerators.
Prior studies showed that the temporal locality of indices in production traces is present among embedding accesses~\cite{recnmp,eisenman2018bandana}.
In Figure~\ref{fig:graph_partition}(a), we propose a locality-aware embedding partition method to identify the hot embedding entries, ranked by the access frequency, to form hot embedding tables.  
The number of entries in the hot embeddings is determined by the capacity budget per thread, \textit{memory capacity / model co-location}.
Hence, the original model graph $G_m$ is partitioned into three sub-graphs: DenseNet $G_d$, SparseNet with full embedding tables $G_s$, and Hot-SparseNet with hot embedding tables $G_{s.hot}$. 
The operator fusion technique~\cite{tvm} is also performed in this stage for element-wise operations.

The partitioned sub-graphs are launched by separate inference threads in a pipelined manner using an intermediate queue as the communication channel.
Mapping the sub-graphs to the hardware devices is model- and hardware-dependent.
{\DesName} considers both the baseline \textit{model-based scheduling} and the partitioned \textit{sparse-dense (S-D) pipeline scheduling}. 
Different model mapping scenarios emerge when applying the two types of scheduling to CPU-only and CPU-Accelerator.

On CPU systems, \textit{model-based scheduling} launches the entire model $G_m$. 
But according to the characterization shown in Sec~\ref{sec:char_host_space}, the operator dependency can incur imbalanced workload allocation among the parallel operator workers.
The idling of operator workers can be reduced by separating SparseNet $G_s$ (no operator dependency) and DenseNet $G_d$ (with operator dependency), which is referred to as \textit{S-D pipeline scheduling}.
In Figure~\ref{fig:graph_partition}(b), parallel operator workers are assigned to each sparse thread to speed up the execution of embedding operations, while a single operator worker is assigned to each dense thread.
Here, communication overhead is taxed to transfer the pooled sparse embedding output.

On CPU-Accelerator, there are three variants derived from the two model scheduling methods.
First, \textit{S-D pipeline scheduling} can be applied by dispatching SparseNet $G_s$ to the host and DenseNet $G_d$ to the accelerator as depicted in Figure~\ref{fig:graph_partition}(c). 
An alternative mapping is shown in Figure~\ref{fig:graph_partition}(d) which dispatches the combined Hot-SparseNet and DenseNet, $G_{s.hot}$+$G_d$, to the accelerator as \textit{model-based scheduling}. 
In this mapping, the host runs the SparseNet $G_s$ threads for the embedding entries excluding $G_{s.hot}$ and then sends the partial sum (Psum) and the remaining embedding indices to the inference threads on the accelerator.
Lastly, to fully utilize the host-side resources, the cores that remain available can perform either \textit{S-D pipeline scheduling} or \textit{model-based scheduling}.

\begin{figure}[t!]
  \centering
  \includegraphics[width=\columnwidth]{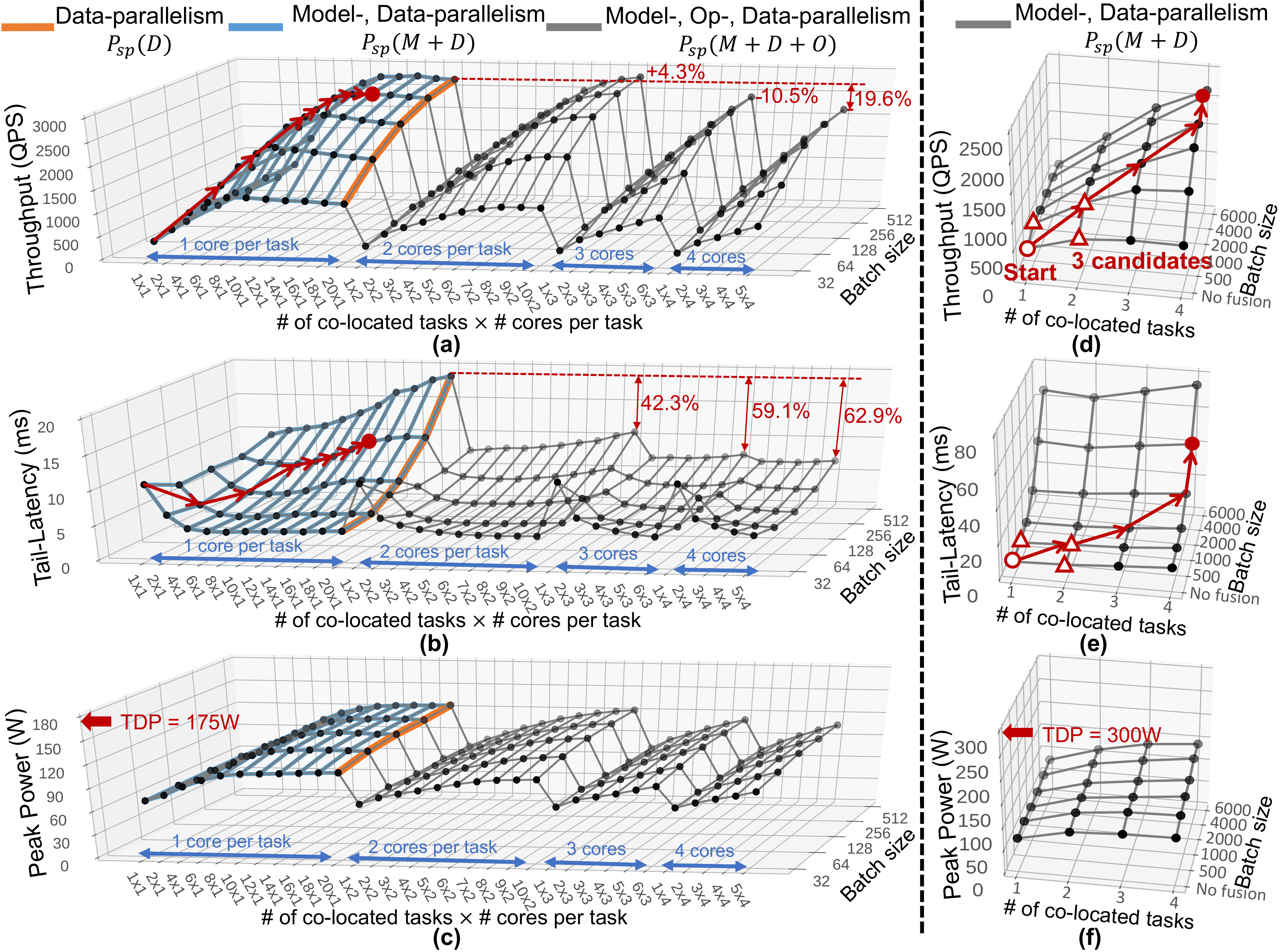}
  \vspace{-0.6cm}
  \caption{Model-based scheduling of DLRM-RMC1 on (a-c) CPU and (d-f) accelerator. 
  Considering all three parallelism dimensions---data, model, and operator---widens the design space, improving latency-bounded throughput and energy efficiency.
  }
  \label{fig:model_exploration}
  \vspace{-0.2cm}
\end{figure}

\textbf{Convexity of Parallelism Space.}
Given every model partition configuration, the task scheduler needs to determine \textit{model-, op- and data-parallelism} in the number of co-located inference threads ($m$), the number of cores allocated per thread ($o$), and the batch sizes ($d$).
On CPUs, the three parallelism spaces are evaluated, $P_{sp}(D)$, $P_{sp}(M+D)$ and $P_{sp}(M+D+O)$ shown in Figure~\ref{fig:model_exploration}(a-c).
$P_{sp}(D)$ is the parallelism space considered in~\cite{DeepRecSys} which only sweeps the batch sizes splitting the large queries and fixes the number of inference threads and the number of cores per thread with $O(d)$ complexity.
$P_{sp}(M+D)$ sweeps the number of inference threads and the batch sizes and fixes the number of cores per thread with $O(m \times d)$ complexity.
$P_{sp}(M+D+O)$ sweeps the number of inference threads, the batch sizes, and the number of cores per thread with $O(m \times d \times o)$ complexity.
Similarly, on the accelerators, shown in Figure~\ref{fig:model_exploration}(d-f), $P_{sp}(M+D)$ sweeps the number of co-located models and the maximum batch sizes fusing queries.
\textit{On both the CPU and the accelerator, the trends of throughput/tail-latency/power in $P_{sp}(M+D)$ are convex.}

\begin{figure}[t!]
  \centering
  \includegraphics[width=\columnwidth]{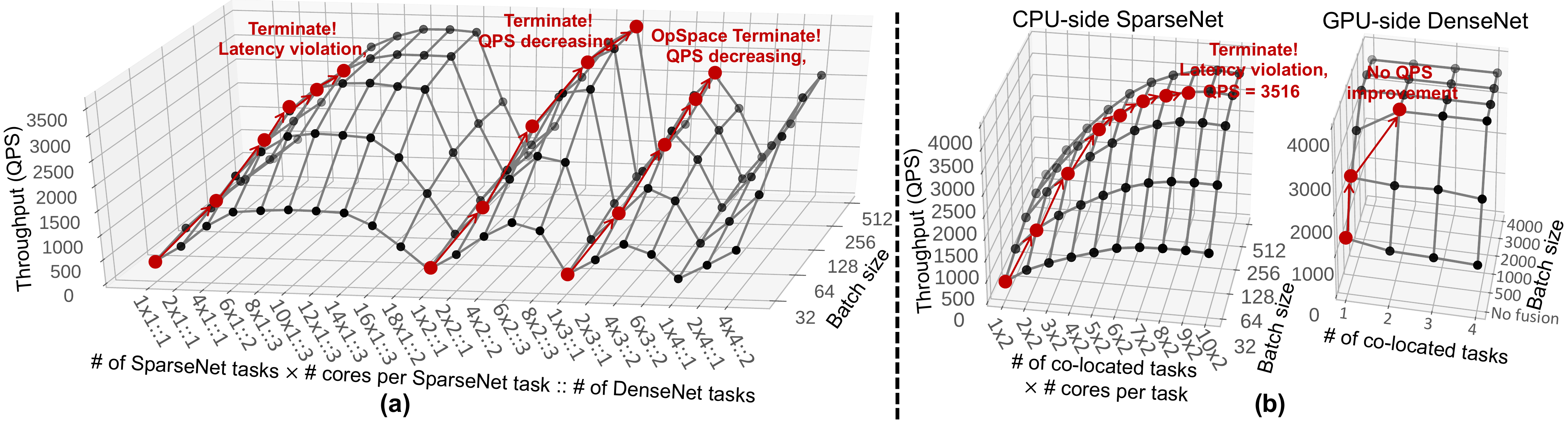}
  \vspace{-0.6cm}
  \caption{Design space of balancing \textit{S-D pipeline scheduling} on (a) CPU and (b) CPU-accelerator platforms.
  }
  \label{fig:DS_exploration}
  \vspace{-0.4cm}
\end{figure}

\begin{algorithm}[!t]
\small
\caption{Gradient-based Search}
\SetAlgoLined

\KwData{SLA latency target $L$,
Provisioned power budget $P$}
\KwResult{Scheduling Configuration}

\textbf{Initialize} $batch\ size$, $num\ of\ threads$\;

\For{(op-parallelism) in $P_{sp}(O)$}{

\# P(M+D) Space Search\;
\While{}{

\# Three candidates\;
$\nabla$Latency[1:3] = Latency$_t$[1:3] $-$ Latency$_{t-1}$\;
$\nabla$QPS[1:3] = QPS$_t$[1:3] $-$ QPS$_{t-1}$\;

\eIf{(max($\nabla$QPS) $>$ 0) and (Latency $<$ L) and (Power $<$ $P$) }{
\# Throughput is increasing\;
update to Configuration(max(Candidate))\;}
{
Terminate search\;
}
}

\If{Configuration[op-parallelism] is decreasing}{
Terminate and return max(Configuration[$P_{sp}(O)$])
}
}
\label{alg:gradient_based_search}

\end{algorithm}

\textbf{Gradient-based Search.}
Considering the convexity observation of $P_{sp}(M+D)$, {\DesName} employs a gradient-based search method to find the optimal parallelism from the exponentially grown search space.
The entire parallelism space, denoted as $P_{sp}(M+D+O)$, combines all the parallelism dimensions and is equal to $P_{sp}(O) \times P_{sp}(M+D)$. 
Here, the operator parallelism space $P_{sp}(O)$ represents the possible allocations of physical CPU cores for one inference thread.
As each core is one operator worker, this allocation can range from one (minimum) to the total number of available physical cores (maximum).
As illustrated in {\em Algorithm~\ref{alg:gradient_based_search}}, given a choice of \textit{op-parallelism} in the set defined by $P_{sp}(O)$, the gradient-based search is performed across the remaining dimensions in $P_{sp}(M+D)$.
It starts from the origin point with minimal model co-location and minimal batch size, as marked by the red hollow dot marked in Figure~\ref{fig:model_exploration}(d). 
At the next step, there are three directions as the candidate configurations, marked by the three red hollow triangles:
(1) increasing the batch size only, 
(2) increasing the number of threads only, and
(3) increasing both the batch size and the number of threads. 
The gradients of tail-latency and throughput of the three candidates relative to the start point are calculated accordingly, represented by the vectors $\nabla$Latency[1:3] and $\nabla$QPS[1:3].
Only candidates that can meet the user-set SLA latency $L$ and power $P$ constraints are considered valid, and {\DesName} picks the one achieving the maximum throughput improvement as the next location to move to. 
Considering the convexity of the $P_{sp}(M+D)$ space, when the throughput results of all three candidates are decreased from the current configuration, the exploration is terminated and the current configuration is reported as the optimum in $P_{sp}(M+D)$ for the current \textit{op-parallelism}.
Finally, the outer-loop that searches over $P_{sp}(O)$ is terminated when the peak throughput of the current \textit{op-parallelism}, $P_{sp}(O)[i]$, starts decreasing as compared with the peak throughput in last \textit{op-parallelism}, $P_{sp}(O)[i-1]$.
In this way, the gradient-based method finds the global optimal configuration achieving the peak latency- and power-bounded throughput across $P_{sp}(M+D+O)$.

Overall, to search the optimal task scheduling configuration, {\DesName} performs the parallelism exploration of $P_{sp}(M+D+O)$ for all possible model partition strategies, include both \textit{model-based scheduling} and \textit{S-D pipeline scheduling}.
Figure~\ref{fig:model_exploration} illustrates how the gradient-based search finds the path to the peak latency-bounded throughput and terminates when the SLA latency is violated for the \textit{model-based scheduling} on multi-core CPU and GPU.
Figure~\ref{fig:DS_exploration} shows the gradient-based search of \textit{S-D pipeline scheduling}.
The gradient-based search determines the allotment for SparseNet and DenseNet to reach S-D pipeline's equilibrium, 
terminated either by SLA latency violation or decreased throughput.
On CPU, as shown in Figure~\ref{fig:DS_exploration}(a), the throughput of each $P_{sp}(M+D)$ space first rises and then falls when sweeping the number of threads divided between SparseNet and DenseNet.
The throughput first climbs up by activating more parallel tasks, increasing both parallelism of SparseNet and DenseNet threads.
Then, the throughput drops with unbalanced pipelining between SparseNet and DenseNet threads.
On CPU-accelerator, as shown in Figure~\ref{fig:DS_exploration}(b), the SparseNet threads are launched on the host-side with $P_{sp}(M+D+O)$ space and DenseNet threads are launched on accelerator with $P_{sp}(M+D)$.
Since the throughput of DenseNet on accelerator is bounded by SparseNet on the host, following each move-step of host-side search, the accelerator-side $P_{sp}(M+D)$ search is performed.
The accelerator-side search is terminated when no more throughput improvement is observed.
The overall search is terminated on both the host and the accelerator upon SLA latency violation.

\subsection{Goal-oriented Cluster Scheduling Optimization}
\label{sec:hyperrec_cluster_schedule}
To minimize the amount of provisioned power avoiding over-provisioning and satisfy the global throughput goal, the heterogeneity-aware cluster scheduler should dynamically allocates the right amount of servers. 
Our characterization in Section~\ref{sec:char_cluster} shows the deficiency in the state-of-the-art greedy scheduler to quantitatively prioritize the allocation of the competing best-matching servers and suggests that a numerical optimization objective is needed to guarantee the global cluster resource cost minimization.

We formulate it as a constrained optimization problem below.
The optimization objective is to find the $N_{1:H,1:M}(t)$ values which minimize the total provisioned power budget shown in Equation (1). 
$N_{h,m}(t)$ is the number of server type $T_h$ assigned to workload $G_m$, and $R$ is the over-provision rate.
$QPS_{h,m}$ and $Power_{h,m}$ are the latency-bounded throughput and provisioned power budget of workload $G_m$ launching on server $T_h$. 
Two constraints need to be satisfied.
First, for all workloads $G_{1:M}$, the number of assigned servers to workload $G_m$ must satisfy the incoming load $load_{m}(t)$ of the workload at time $t$ in Equation (2).
Second, the number of activated servers should not exceed the capacity limit in Equation (3), where $N_{h}$ is the amount of available $T_h$ servers.

\vspace{-0.2cm}
\[Minimize \sum_{m=1}^{M}(\sum_{h=1}^{H}(N_{h,m}(t)\times Power_{h,m}))\ (1),\ subject\ to\]

\vspace{-0.2cm}
\[\forall m\in [1..M], \sum_{h=1}^{H}(N_{h,m}(t)\times QPS_{h,m}) \geqslant load_{m}(t)(1+R\%)\ (2),\]

\vspace{-0.2cm}
\[\forall h\in [1..H], \sum_{m=1}^{M}N_{h,m}(t) \leqslant N_{h}\ \ (3)\]

In Figure~\ref{fig:design_flow}(c), the cluster manager keeps the \textit{efficiency tuple} of the latency-bounded QPS and the provisioned power for every server $T_h$ and workload $G_m$ pair, and knows $load_{1:M}(t)$ for all workloads at the current time $t$.
After launching the optimization solver getting $N_{1:H,1:M}(t)$, the cluster manager can activate/release servers and decide which workload launched on the activated servers.
The dynamic provision is performed at coarse time-interval (10s of minutes) to amortize the overhead of workload setup time (10s of seconds).
The over-provision rate $R$ is set to handle the load increment during this time-interval. $R$ is estimated by profiling history loads changes during the length of time-interval.
To solve the linear constraint problem, many standard solution have been proposed, e.g., simplex and interior-point~\cite{interior-point}.

\section{Experimental Methodology}
\label{sec:method}

\begin{figure}[t!]
  \centering
  \includegraphics[width=\columnwidth]{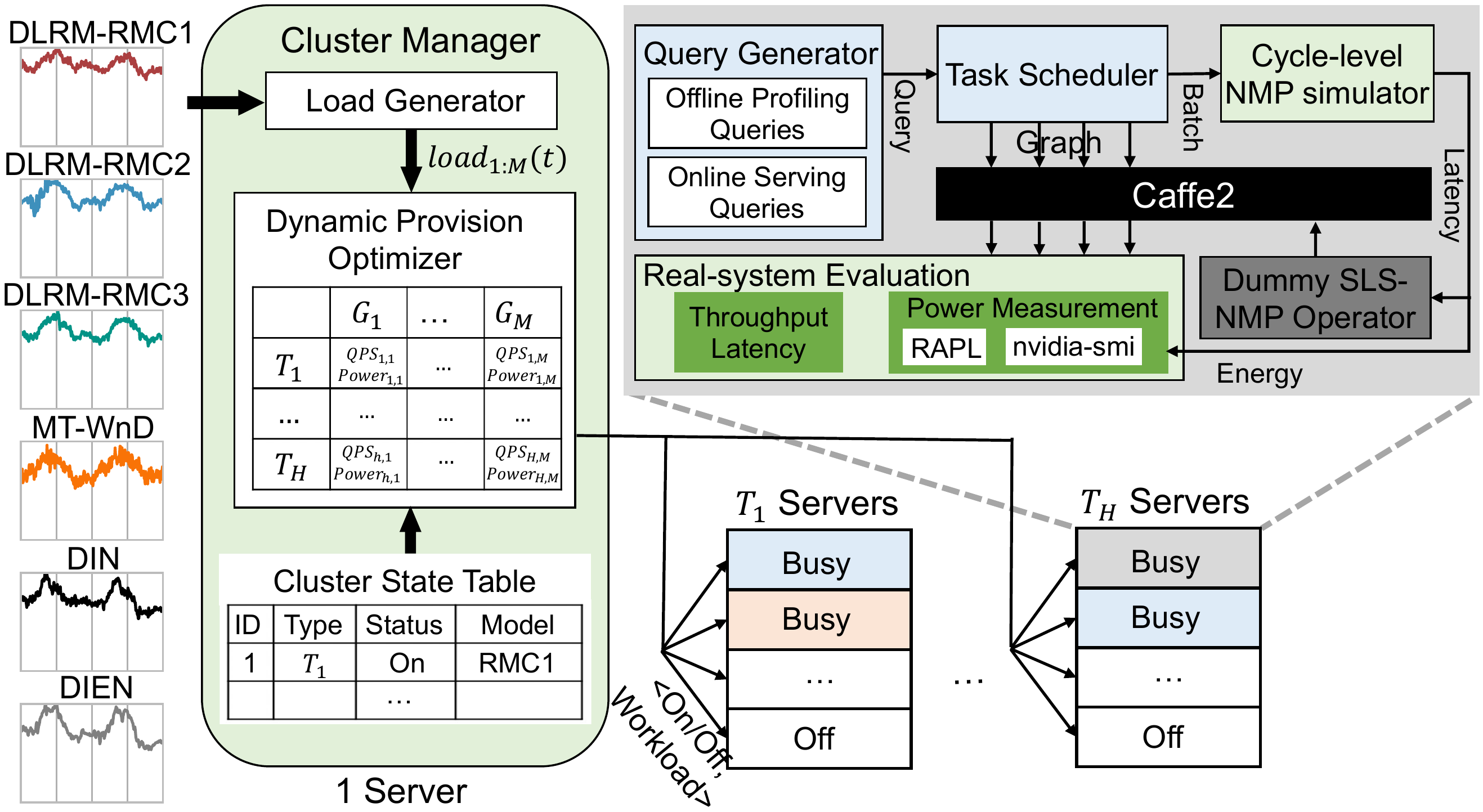}
  \vspace{-0.6cm}
  \caption{Implemented cluster prototype including one cluster manager server with load generator, dynamic provision optimizer and cluster state table, and the different $T_1$--$T_H$ servers given the availability constraints $N_1$--$N_H$.}
  \label{fig:cluster_test}
  \vspace{-0.6cm}
\end{figure}

\textbf{{\DesName} Implementation.} 
We implement a prototype cluster as depicted in Figure~\ref{fig:cluster_test}.
One server in the cluster is used as the cluster manager to determine server activation or release. We use a trace-driven load generator to generate inference requests following the query arrival characteristics observed in production for the $M$ workloads, $load_{1:M}(t)$ (Section~\ref{sec:hyperrec_cluster_schedule}).
The dynamic provision optimizer maintains a lookup table for provisioned power ($Power_{1:H,1:M}$) and achievable throughput ($QPS_{1:H, 1:M}$), and a cluster state table tracks the status of servers in the cluster.
The cluster manager runs an optimizer program that uses an interior-point solver~\cite{interior-point} to obtain the optimal allocation solution, $N_{1:H,1:M}(t)$.
Based on the cluster state, the cluster manager identifies which servers should be activated/released and which workload to launch.
Each server in the cluster launches the {\DesName} task scheduler extended from DeepRecSys~\cite{DeepRecSys}.
The query generator loads the inference queries and feeds them to the task scheduler.
It then launches the inference tasks written in Caffe2~\cite{caffe2} on CPUs and GPUs.

\textbf{System Configuration.} 
In our prototype cluster, we have a limited number of available $T_1$--$T_{10}$ servers, denoted as $N_1$--$N_{10}$.
Table~\ref{tab:sys_config} summarizes the $T_1$--$T_{10}$ system configurations.
The various server types are constructed with a permutation of different combinations of \textit{CPU}+\textit{memory}+\textit{GPU}.
They represent the system heterogeneity at modern data centers.
We include two generations of Intel Xeon processors: CPU-T1 and CPU-T2 
and two types of memory DIMMs: DDR4 and a DIMM-based NMP solution~\cite{recnmp}.
NMP$\times N$ represents $N$ ranks in one memory channel to attain $N\times$ rank-level parallelism.
We also use two generations of NV GPUs, P100 and V100, as DL accelerators.
The NMPs and GPUs are selected to represent diverse memory-centric and compute-centric hardware acceleration approaches.

\textbf{Evaluation Framework.} 
All evaluations of server types consisting only of CPU and GPU with DDR4 are performed on real systems.
To quantify the benefits provided by the memory-centric NMP solutions, we follow the emulation-based methodology adopted by earlier work~\cite{tensordimm,recnmp,Tensorcasting} that combines the performance of real systems (e.g. CPU/GPU) with a cycle-level NMP simulator~\cite{recnmp}.
The cycle-level NMP simulation of sampled inference queries is performed in advance to record the embedding operators' latency and energy in a lookup table (LUT).
Thus, the time-consuming simulation is avoided during real-system evaluation to keep up with the real execution.
The dummy SLS-NMP operator is launched by setting the pooling factor to 1 and taxing the latency from the LUT for the current batch's embedding operation.
Its energy cost is sent to the power measurement module.
Throughput, tail-latency, and power consumption are captured by the performance and power monitor modules.
These power measurements account for all the system components.
CPU and DDR4 powers are read from Intel RAPL~\cite{RAPL}, and GPU power is measured by Nvidia API \textit{nvidia-smi}.
These server-level performance and power measurement can be extrapolated for rack-level study based on the cluster composition.

\begin{table}[t!]
\scriptsize

\caption{System Parameters and Configurations}
\vspace{-0.2cm}

\label{tab:sys_config}
\centering
\begin{tabular}{|c|c|c|c||c|c|c|c|c|}
\hline
$T_h$ & $N_h$ & CPU    & Memory & $T_h$ & $N_h$ & CPU    & Memory & GPU  \\ \hline
$T_1$    & 100 & CPU-T1 & DDR4   & $T_6$  & 10   & CPU-T1 & DDR4   & P100 \\ \hline
$T_2$    & 100 & CPU-T2 & DDR4   & $T_7$  & 5   & CPU-T2 & DDR4   & V100 \\ \hline
$T_3$    & 15 & CPU-T2 & NMPx2  & $T_8$  & 6   & CPU-T2 & NMPx2  & V100 \\ \hline
$T_4$    & 10 & CPU-T2 & NMPx4  & $T_9$  & 4   & CPU-T2 & NMPx4  & V100 \\ \hline
$T_5$    & 5 & CPU-T2 & NMPx8  & $T_{10}$  & 2  & CPU-T2 & NMPx8  & V100 \\ \hline
\end{tabular}

\vspace{0.1cm}

\begin{tabular}{|c|c|c|}
\hline
CPU Server & CPU-T1 & CPU-T2 \\ \hline
Chip & Intel Xeon D-2191 & Intel Xeon Gold 6138 \\ \hline
Frequency & 1.6 GHz & 2.0 GHz \\ \hline
Physical Cores & 18 & 20 \\ \hline
L1/L2 size & \multicolumn{2}{c|}{32 KB / 1 MB} \\ \hline
LLC size & 24.75 MB & 27.5 MB \\ \hline
TDP & 86 W & 125 W \\ \hline
\end{tabular}

\vspace{0.1cm}

\begin{tabular}{|c|c|c|c|c|c|}
\hline
\multirow{2}{*}{Memory} & \begin{tabular}[c]{@{}c@{}}DDR4\\ (CPU-T1)\end{tabular} & \begin{tabular}[c]{@{}c@{}}DDR4\\ (CPU-T2)\end{tabular} & \begin{tabular}[c]{@{}c@{}}NMP\\ x2\end{tabular} & \begin{tabular}[c]{@{}c@{}}NMP\\ x4\end{tabular} & \begin{tabular}[c]{@{}c@{}}NMP\\ x8\end{tabular} \\ \cline{2-6} 
                  & \multicolumn{2}{c|}{Real system} & \multicolumn{3}{c|}{Simulation} \\ \hline
Memory Channels   & 4              & 4               & 4         & 4        & 4        \\ \hline
DIMM per Channel & 1              & 1               & 1         & 2        & 4        \\ \hline
Ranks per DIMM    & 1              & 2               & 2         & 2        & 2        \\ \hline
Capacity (GB)     & 64          & 128               & 128        & 256      & 512   \\ \hline
TDP (Watt)        & 28           & 50               & 50         & 100      & 200    \\ \hline
\end{tabular}

\vspace{0.1cm}

\begin{tabular}{|c|c|c|}
\hline
GPU             & Nvidia P100          & Nvidia V100        \\ \hline
GPU Boost Clock & 1480 MHz             & 1530 MHz           \\ \hline
SMs / TPCs      & 56 / 28              & 80 / 40            \\ \hline
Memory          & \multicolumn{2}{c|}{16 GB HBM @ 900 GB/s} \\ \hline
Interface       & \multicolumn{2}{c|}{PCIe Gen3 @ 16 GB/s}  \\ \hline
TDP             & \multicolumn{2}{c|}{300 W}                \\ \hline
\end{tabular}

\vspace{-0.4cm}
\end{table}
\section{Performance Evaluation}
\label{sec:evaluation}
In this section, we analyze various benefits and implications from applying our {\DesName} framework across the industry-representative models and heterogeneous server types. 
We have performed a comprehensive evaluation at both the server and cluster levels. All evaluation is performed with the production-scale model size using the locality-aware model partition to satisfy the memory capacity constraint of the NV P100 and V100 accelerators.

\subsection{Task Scheduling Exploration}
{\DesName} identifies the under-explored parallelism space of task scheduling beyond the state-of-the-art baseline~\cite{DeepRecSys,Baymax}.
{\DesName} considers the different model partition strategies and explores the parallelism space $P_{sp}(M+D+O)$ on the CPU and $P_{sp}(M+D)$ on the accelerator.
The baseline task scheduler considered here applies \textit{model-based scheduling} on both the CPU and the accelerator, then explores $P_{sp}(D)$ on the CPU (DeepRecSys~\cite{DeepRecSys}) and \textit{model co-location} on the accelerator (Baymax~\cite{Baymax}).
With the thorough exploration of these expanded spaces, the {\DesName} task scheduler achieves significant performance improvement.

\begin{figure}[t!]
  \centering
  \includegraphics[width=\columnwidth]{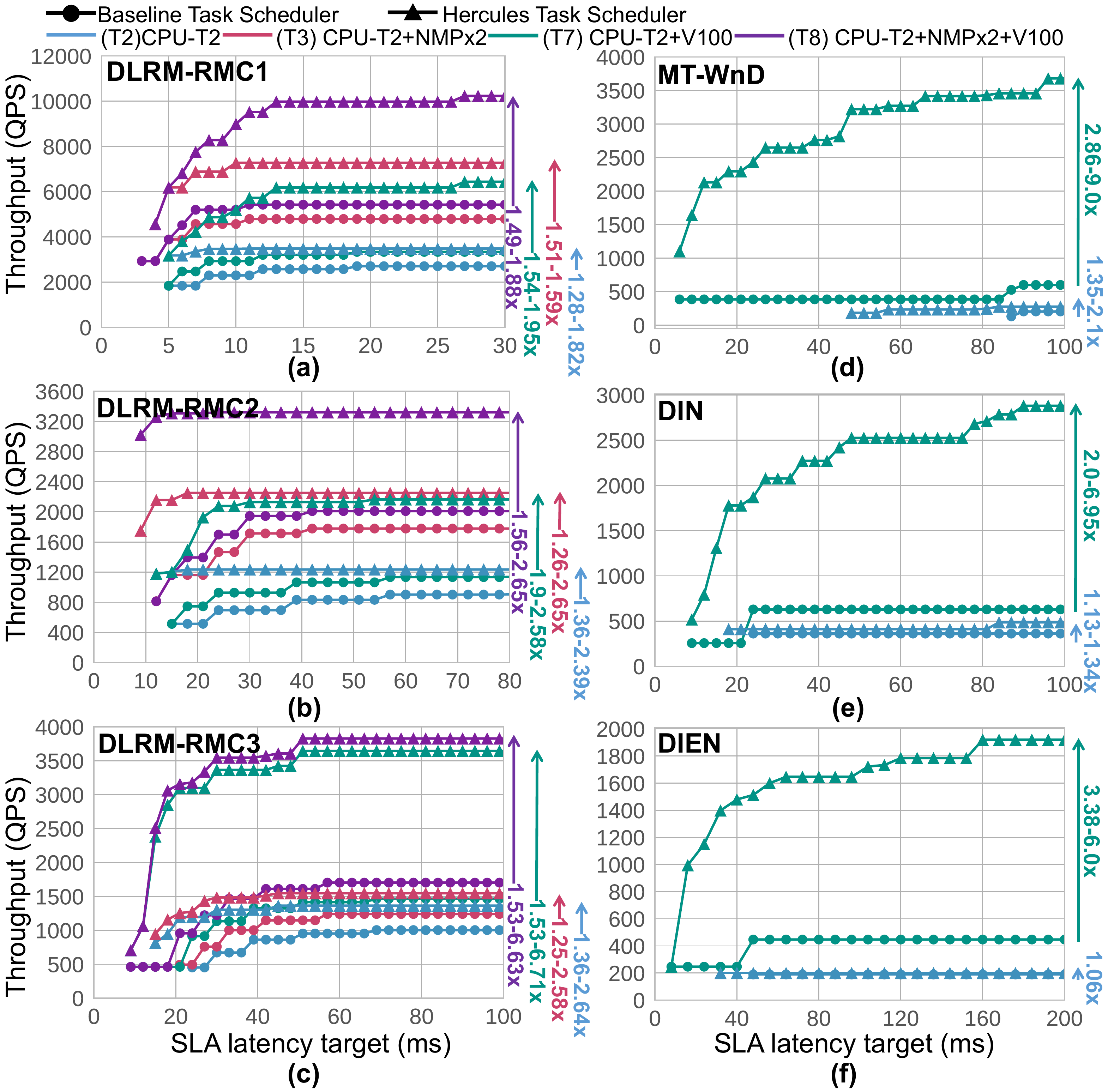}
  \vspace{-0.6cm}
  \caption{
  Comparison of SLA-aware task schedulers, the baseline (DeepRecSys~\cite{DeepRecSys} on the CPU and Baymax~\cite{Baymax} on the accelerator) and {\DesName}, for (a) DLRM-RMC1, (b) RMC2, (c) RMC3, (d) MT-WnD, (e) DIN, (f) DIEN.
  }
  \label{fig:model_vs_ds}
  \vspace{-0.4cm}
\end{figure}

Figure~\ref{fig:model_vs_ds} shows the latency-bounded throughput of the six models with {\DesName} and the baseline scheduler.
On the CPU-centric (CPU-only and CPU+NMP) servers without GPU, the combination of \textit{S-D pipeline scheduling} and the thorough exploration of $P_{sp}(M+D+O)$ improves the latency-bounded throughput of models with multi-hot embedding operations. DLRM-RMC1/RMC2/RMC3 are accelerated by up to 1.82$\times$/2.39$\times$/2.64$\times$ on server $T_{2}$ (CPU-only) and 1.59$\times$/2.65$\times$/2.58$\times$ on server $T_{3}$ (CPU+NMP), respectively.
For MT-WnD/DIN/DIEN where SparseNet contributes to less than 5\% of the end-to-end latency, the improvement from the task scheduler is lower.
On CPU+GPU servers, the thorough exploration of $P_{sp}(M+D)$ space considering model co-location and query fusion improves the latency-bounded throughput substantially for compute-dominated models. For example, DLRM-RMC3/MT-WnD/DIN/DIEN are accelerated by up to 6.71$\times$/9.0$\times$/6.95$\times$/6.0$\times$ on server $T_{7}$ (CPU+GPU).

\subsection{Server Architecture Exploration}
Tailoring the server architecture for the workloads is one major approach to improve cluster efficiency.
However, given the rather large design space, it can be an onerous task to optimally match an appropriate high-performance and energy-efficient server architecture to a workload for cluster capacity and provisioned power savings.
{\DesName}'s offline profiling performs the server architecture exploration.
We consider the hardware design space consisting of 10 server architectures represented as $T_1$--$T_{10}$ in Table~\ref{tab:sys_config}.
In which, $T_{1}$--$T_{2}$ (CPU-only) and $T_{6}$--$T_{7}$ (CPU+GPU) are generally applied to all six workloads.
The NMP-enabled servers, $T_{3}$-$T_{5}$ and $T_{8}$-$T_{10}$, target the multi-hot encoded embedding operations with Gather-Reduce pattern in all three DLRM models; nonetheless, they do not show as much benefit for MT-WnD, DIN and DIEN which perform embedding lookup only with no pooling operation.

\begin{figure*}[t!]
  \centering
  \includegraphics[width=\textwidth]{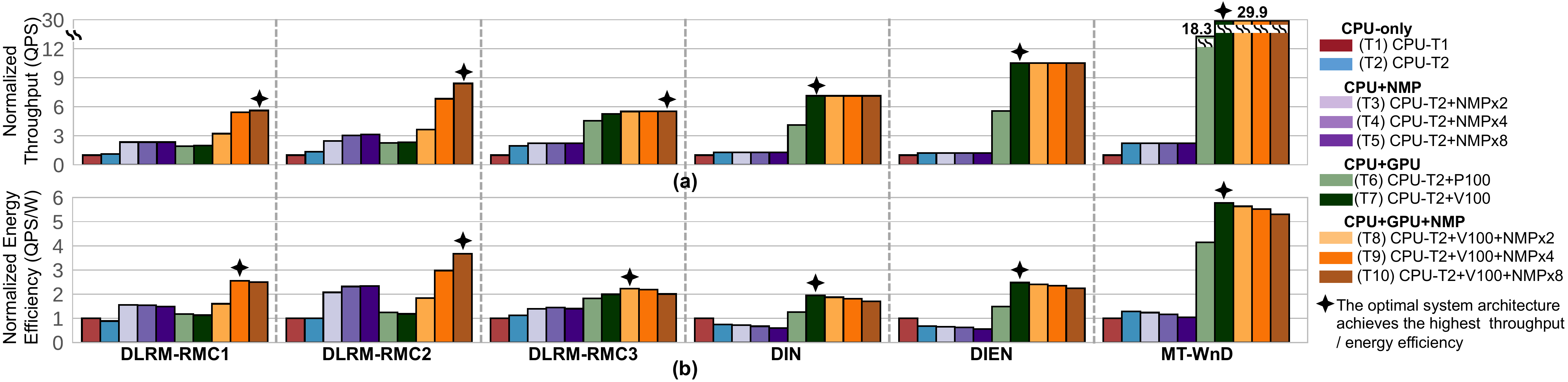}
  \vspace{-0.6cm}
  \caption{Normalized latency-bounded (a) throughput (QPS) and (b) energy efficiency (QPS-per-Watt) of DLRM-RMC1, RMC2, RMC3, DIN, DIEN and MT-WnD with 20ms, 50ms, 50ms, 50ms, 100ms, 100ms as the SLA latency target on the $T_1$--$T_{10}$ different server architectures.}
  \label{fig:hw_efficiency}
\end{figure*}

\begin{figure*}[t!]
\begin{minipage}[t]{0.40\linewidth}
    \includegraphics[width=\linewidth]{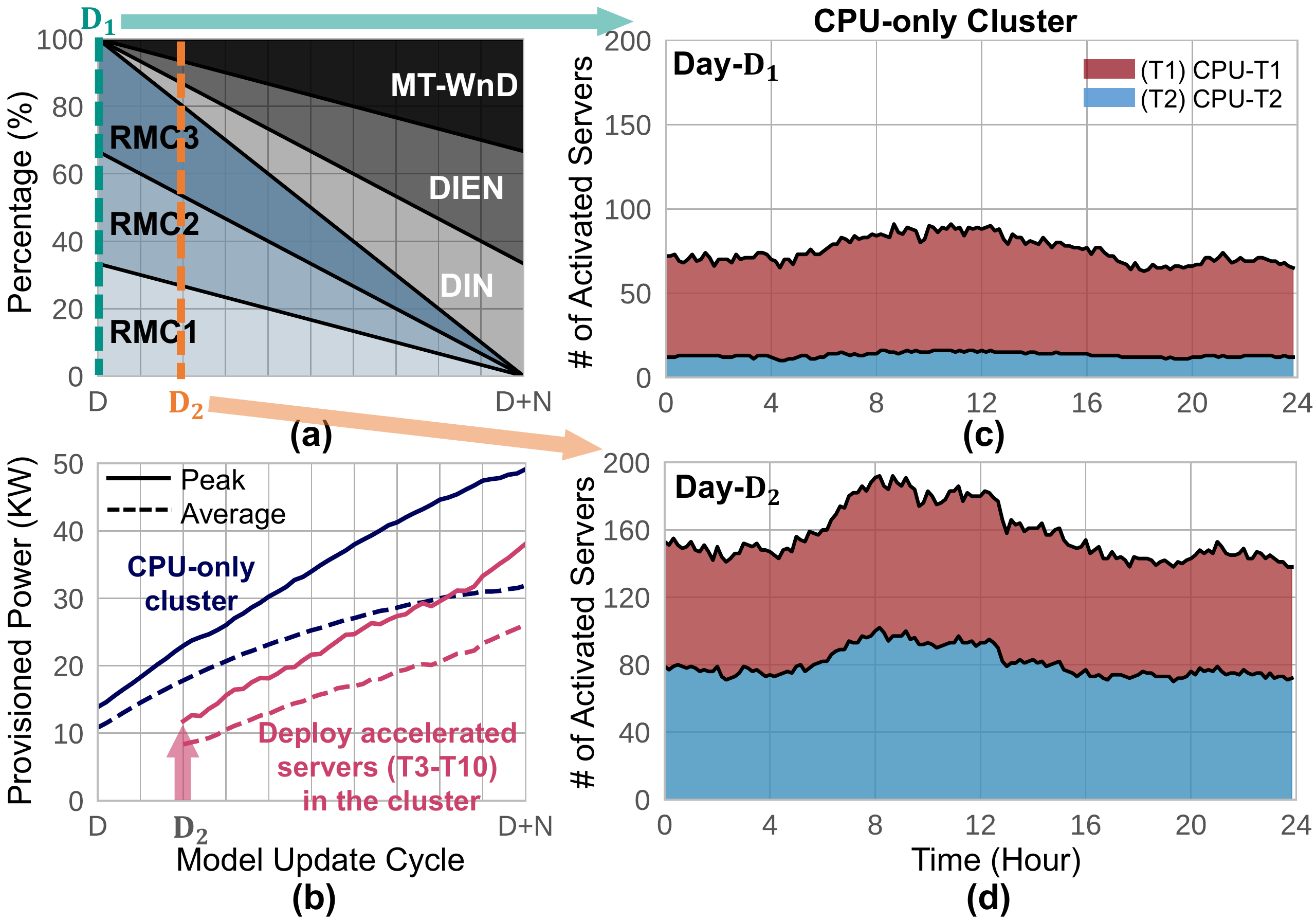}
    \vspace{-0.7cm}
    \caption{(a) Synthetic model evolution; 
    (b) Peak and average provisioned power during evolution.
    Capacity provisioning of CPU-only cluster on (c) Day-$D_1$ and (d) Day-$D_2$.}
    \label{fig:model_evlove}
\end{minipage}%
    \hfill%
\begin{minipage}[t]{0.58\linewidth}
    \includegraphics[width=\linewidth]{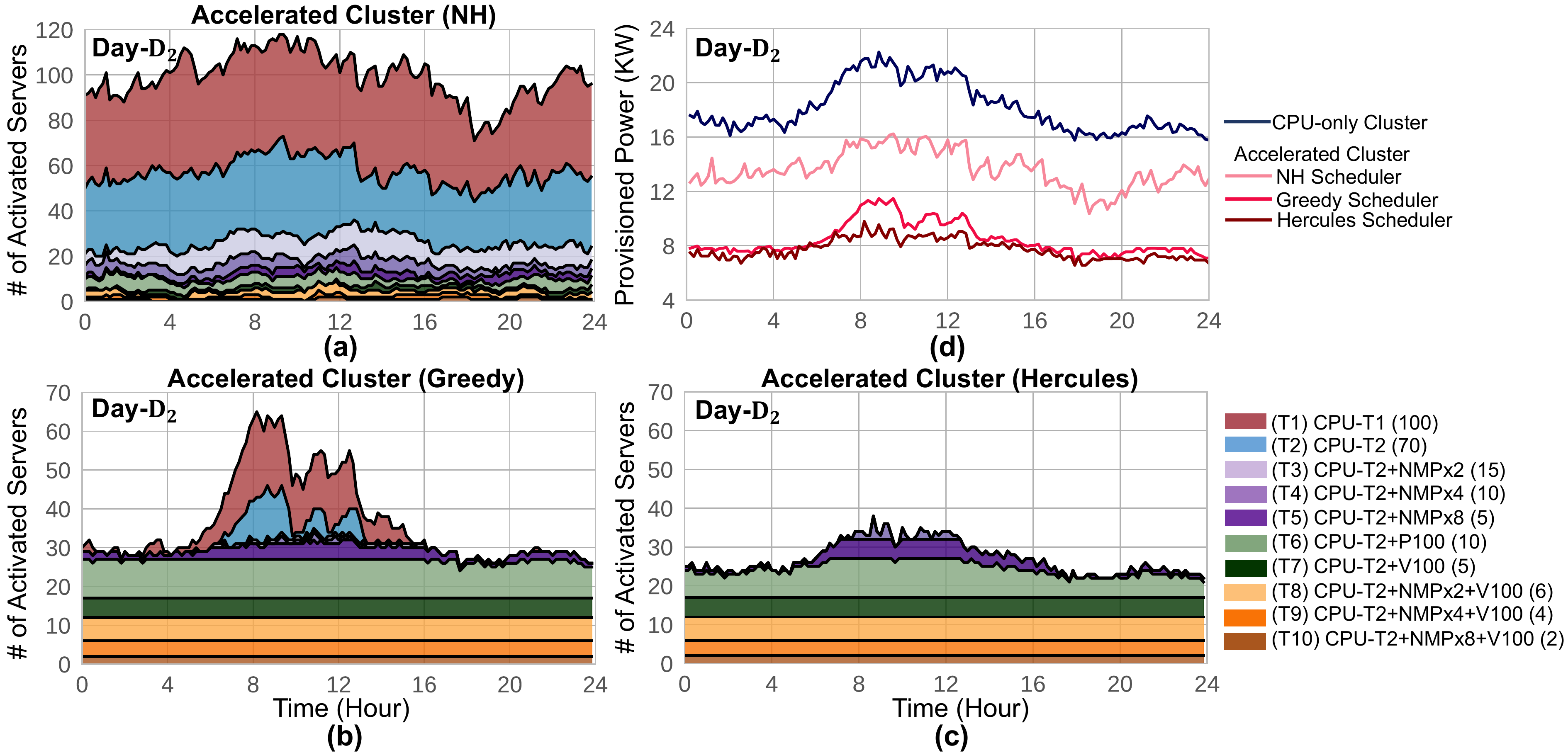}
    \vspace{-0.7cm}
    \caption{Cluster capacity provisioning of accelerated cluster with (a) NH scheduler, (b) greedy scheduler and (c) {\DesName} scheduler. 
    (d) The comparison of provisioned power budget.}
    \label{fig:hardware_acceleration}
\end{minipage} 

\vspace{-0.6cm}
\end{figure*}

Our results indicate that the optimal server architecture varies across the space of workloads and system configurations.
As shown in Figure~\ref{fig:hw_efficiency}, {\DesName} evaluates across 6$\times$10 workload-server pairs in term of latency-bounded throughput (QPS) and energy efficiency (QPS-per-Watt) with user SLA targets.
For the CPU-only servers, server $T_2$ achieves higher throughput across all six workloads over server $T_1$, since CPU-T2 has more physical cores and higher frequency. 
However, it comes with a cost of higher power, too.
Measured by QPS-per-Watt, server $T_1$ is shown to be more energy-efficient for DLRM-RMC2, DLRM-RMC3, and MT-WnD.

Using GPU as the accelerator, the CPU+GPU servers $T_6$ and $T_7$ achieve significant throughput and energy efficiency improvement over the CPU-only server $T_2$ for the compute-dominated workloads such as DLRM-RMC3, MT-WnD, DIN, and DIEN. 
Note that energy efficiency improvement of the CPU+GPU servers is constrained by GPUs' high leakage power and the upper bound of batch size to meet the SLA latency target.

With near-memory acceleration, $T_3$--$T_5$ and $T_8$--$T_{10}$ servers achieve significant throughput and energy efficiency improvement for memory-dominated DLRM-RMC1 and RMC2 over $T_2$ (CPU-only) and $T_7$ (CPU+GPU) baselines. 
The improvements become smaller for DLRM-RMC3 which is compute-dominated. 
For DIN, DIEN, and MT-WnD models, the energy efficiency improvement is even lower with no throughput improvement.
With only one-hot embedding lookup operations in DIN, DIEN and MT-WnD that have no Gather-Reduce operations on the memory side,
NMP-DIMMs behave exactly the same as regular DRAM DIMMs (similar throughput).
However, the NMP$\times$2, NMP$\times$4 and NMP$\times$8 configurations (Table~\ref{tab:sys_config}) dissipate extra idle power for NMP processing units and more number of DIMMs (lower energy efficiency) than the DDR4 configuration.

\subsection{Cluster Heterogeneity-aware Provision}
Finally, we evaluate the cluster manager during the online serving phase to efficiently and dynamically allocate workloads on the right amount of best-matching servers.

\textbf{Model Evolution.}
In this experiment, we mimic model evolution by varying the composition of the workload sets. 
In Figure~\ref{fig:model_evlove}(a), we assume the synthetic evolution process is linear.
The recommendation workloads, initially consisting of DLRM-RMC1, RMC2 and RMC3, are gradually replaced by DIN, DIEN and MT-WnD which represent new models with higher accuracy and increased complexity.
In the cluster with CPU-only servers ($T_1$ and $T_2$), the increasing ratio of the new models requires more cluster capacity and provisioned power budget.
Comparing between the one-day snapshots of Day-$D_2$ and Day-$D_1$, 20\% of the incoming loads are routed to the higher complexity models.
The cluster capacity (number of activated servers) in Figure~\ref{fig:model_evlove}(c)(d) and the provisioned power in Figure~\ref{fig:model_evlove}(b) are increased by 2.27$\times$ and 1.77$\times$ at peak and 2.09$\times$ and 1.64$\times$ on average.
So, with only the baseline CPU servers deployed, by the end of model evolution, the cluster capacity and provisioned power are projected to increase 5.4$\times$ and 3.54$\times$.

\textbf{Comparison with Prior Cluster Scheduler.}
Deploying hardware accelerators in the cluster can improve cluster execution efficiency, but it also brings an increasing level of system heterogeneity.
In practical datacenter deployment, the number of available accelerated servers are usually limited.
We assume the accelerated servers $T_{3}$-$T_{10}$ with the limited amount $N_{3}$-$N_{10}$ (15, 10, 5, 10, 5, 6, 4, 2) are deployed on Day-$D_2$.
In Figure~\ref{fig:model_evlove}(b), the accelerated cluster achieves significant improvement, 22--52\% and 18--54\% of peak and average provisioned power, during the model evolution process.

The cluster scheduler determines the priority of the server allocation for the workloads.
We evaluate the three cluster schedulers, heterogeneity-oblivious (NH) scheduler, greedy scheduler~\cite{paragon,Quasar}, and {\DesName} scheduler.
Figure~\ref{fig:hardware_acceleration} shows cluster capacity and power provisioning of the three cluster schedulers for the one-day snapshot of Day-$D_2$.
Both greedy scheduler and {\DesName} scheduler achieve high utilization of the accelerated servers in the cluster.
Compared with the NH scheduler which ignores the hardware heterogeneity, the greedy scheduler is heterogeneous-aware and prioritizes workloads allocated to the optimal available servers.
The greedy scheduler achieves capacity saving by 75.8\% (peak) and 67.4\% (average) and provisioned power saving by 50.8\% (peak) and 42.7\% (average) over the NH scheduler.
Our {\DesName} scheduler rigorously formulates the global optimization objectives to quantitatively prioritize server allocation for cluster resource cost minimization and is able to further save the cluster capacity by 47.7\% (peak) and 22.8\% (average) and the provisioned power by 23.7\% (peak) and 9.1\% (average) over the greedy scheduler.

\section{Conclusion}
As production-grade recommendation systems continue to demand more datacenter resources, optimizing their serving performance and efficiency is important for the overall infrastructure cost.
In this paper, we perform an in-depth characterization of the state-of-the-art recommendation serving framework to identify system inefficiency.
We propose {\DesName}, a framework to efficiently serve recommendation inference 
using a two-stage optimization procedure: offline profiling and online serving. {\DesName} explores the task scheduling space for individual servers and dynamically provisions the best-matched heterogeneous datacenter resources in response to real-time load fluctuations.
{\DesName}'s task scheduling achieves 1.03$\times$ to 9$\times$ latency-bounded throughput improvement for individual servers while the heterogeneity-aware cluster manager saves up to 47.7\% and 23.7\% cluster capacity and provisioned power, respectively, over the state-of-the-art greedy scheduler.

\section*{Acknowledgments}
The authors would like to thank the anonymous reviewers for their valuable comments and suggestions.
Liu Ke and Xuan Zhang were partially supported by NSF CCF-1942900.

\renewcommand{\baselinestretch}{0.935}
\bibliographystyle{IEEEtranS}
\let\OLDthebibliography\thebibliography
\renewcommand\thebibliography[1]{
  \OLDthebibliography{#1}
  \setlength{\itemsep}{3pt}
}
\bibliography{main}

\end{document}